\documentclass[onecolumn,nobibnotes]{revtex4}

\usepackage{amsmath}
\usepackage{graphicx}
\usepackage{float}
\usepackage{subfigure}
\usepackage{wrapfig}
\usepackage{multirow}
\usepackage{tabularx}
\usepackage{array}
\usepackage{url}
\usepackage{tikz}
\usepackage{comment}
\usetikzlibrary{positioning}
\usepackage{graphicx,color}

\usepackage{hyperref}
\hypersetup{
    colorlinks=true,
    linkcolor=blue,
    citecolor=blue,
    }

\def\lsim{\raise0.3ex\hbox{$<$\kern-0.75em\raise-1.1ex\hbox{$\sim$}}}
\def\gsim{\raise0.3ex\hbox{$>$\kern-0.75em\raise-1.1ex\hbox{$\sim$}}}

\def\beq{\begin{equation}}
\def\eeq{\end{equation}}
\def\beqa{\begin{eqnarray}}
\def\eeqa{\end{eqnarray}}

\def\gappeq{\mathrel{\rlap {\raise.5ex\hbox{$>$}}
{\lower.5ex\hbox{$\sim$}}}}

\def\lappeq{\mathrel{\rlap{\raise.5ex\hbox{$<$}}
{\lower.5ex\hbox{$\sim$}}}}

\def\Toprel#1\over#2{\mathrel{\mathop{#2}\limits^{#1}}}

\begin{document}

\title{Initial-state geometry and multiplicity distributions in pp and pPb 
collisions}

\author{R. Terra$^1$, A. V. Giannini$^{2,3}$ and F. S. Navarra$^1$ }
\affiliation{$^1$Instituto de F\'{\i}sica, Universidade de S\~{a}o Paulo, 
Rua do Mat\~ao, 1371, CEP 05508-090,  S\~{a}o Paulo, SP, Brazil\\
$^2$ Faculdade de Ciências Exatas e Tecnologia, Universidade Federal da 
Grande Dourados, CEP 79804-970,  Dourados, MS, Brazil\\
$^3$ Departamento de F\'isica, Universidade do Estado de Santa Catarina, 
89219-710 Joinville, SC, Brazil
}

\begin{abstract}
This work investigates the possibility of accessing the initial geometric 
shape of the proton in proton-proton and proton-nucleus collisions at the 
LHC. In particular, we look for manifestations of the configuration in which 
the proton is made of three 
quarks linked by a Y-shape gluon string, called baryon junction. This     
initial state spatial configuration has been used in the past to describe  
data on baryon rapidity distributions, diffractive $J/\psi$ production and 
multiplicity distributions in pp collisions. In spite of its success in 
explaining the data, the evidence of the baryon junction still needs  
confirmation. Further studies will be undertaken at the electron-ion collider. 
In this work we study multiplicity distributions measured in pp and pPb     
collisions. Different initial state geometries are used as input in a Monte  
Carlo event generator which implements the $k_T$-factorization formalism of 
the CGC with KLN unintegrated gluon distributions. The results show that the 
data on multiplicity distributions are good enough to discriminate between 
different initial state geometries. Moreover, they indicate that it is crucial  
to take into account the intrinsic fluctuations of the saturation scale.

\end{abstract}
\maketitle

\section{Introduction}

\subsection{The baryon junction} 

Despite all the progress, we are still far from a complete picture of the 
nucleon. Determining the geometric structure of the proton is one of the  
great challenges in high energy physics. It is well established that the  
proton consists of quarks and gluons, collectively called partons and  
it is important to understand how these partons are distributed within the 
proton. 

Long ago \cite{venez}, Rossi and Veneziano suggested that the nucleon is 
made by three  quarks located in the vertices of a Y-shaped gluon string, 
or baryon junction (BJ).  The intermediate (Fermat) point was introduced in 
order to guarantee the  gauge invariance of the baryon wave function. 
This picture received support later, when Y-shaped flux tubes were found in 
quenched lattice QCD simulations \cite{latt1,latt2,latt3,latt4} of a system 
of three static quarks.  

The BJ picture of the nucleon was brought to the context of high energy  
proton-proton (pp) collisions by D. Kharzeev  in \cite{dima}, where  he 
suggested that the BJ would be responsible for a new mechanism of baryon   
stopping. In the standard picture of leading baryon formation \cite{fowler},  
the valence  quarks recombine after the collision, forming a  baryon at large 
rapidities. In contrast, in \cite{dima} it was conjectured that the baryon     
junction breaking could lead to a low rapidity (``stopped'') baryon plus three 
mesons with large rapidities.  
The idea was  implemented in analytical models \cite{bopp06} and also in the 
HIJING/B Monte-Carlo event generator \cite{top04}. The predictions of these 
models were in agreement with the available experimental data
\cite{stopping1,stopping2,stopping3,stopping4,stopping5}. 

Later the BJ was used in \cite{schenke} as initial condition in the 
study of coherent and incoherent diffractive production of $J/\psi$ in 
electron-proton scattering with the IP-Sat and IP-Glasma models. The authors 
compared the results obtained with the BJ (which they called ``stringy'' 
proton)  with the results obtained with three constituent quarks of finite 
size as initial condition.  The conclusion was that, even though using the 
BJ initial condition leads to good agreement with data, the precise shape of 
the proton could not be constrained by incoherent diffractive $J/\psi$ production. 
They also concluded that the data were well reproduced when they considered 
event-by-event geometric fluctuations, as well as the intrinsic  saturation 
scale fluctuations as done in \cite{larry16}.

Another evidence of the baryon-junction was reported in \cite{nosso}, where  
the authors found that the BJ is able to explain the accelerated growth 
of the $J/\psi$ yield  observed in high multiplicity events in pp 
collisions.

All these phenomenological successes are not enough to unambiguously prove the 
existence of the baryon junction because all the measurements quoted above could 
still be explained with other models. Besides, one could argue that, increasing 
the resolution scale ($Q^2$) or boosting the nucleon to higher energies 
(decreasing $x$) the simple geometrical picture implied by the BJ is blurred
by the profusion of produced partons during the DGLAP and/or BFKL evolution.

Fortunately, there are good reasons to believe that the Y-shape configuration
is not washed out and manifests itself even in high energy collisions. Indeed, 
there are indications that the geometric organization of matter persists even 
at LHC energies. In \cite{manty} it was shown that exclusive vector meson 
production is sensitive to the geometric deformation of the target. The     
authors concluded that the nuclear geometry (and its deformations)  
is not washed out by small-x evolution and  
hence the deformations previously inferred from low-energy experiments will 
be visible in high-energy collisions. This suggests  the Y shape of the    
proton could survive the quantum evolution and manifest itself in high energy 
pp collisions. In view of the uncertainties involving this issue,  
more calculations and more data are needed. Along this direction, there is an 
interesting proposal to look for the BJ at the forthcoming electron-ion 
collider \cite{bran22}.
 
\subsection{The baryon junction and multiplicity distributions}

In this work we look for manifestations of the baryon junction in multiplicity 
distributions of charged particles measured  at the LHC. Multiplicity is a 
global observable that allows us to characterize events in all colliding systems 
and has been  studied as a tool to understand multiparticle production 
processes. From the experimental point of view, charged-particle multiplicity
is one of the simplest observables and it allows us to gain insights on the 
production mechanisms \cite{mult1,mult2}.

In high energy proton (nucleus) - proton (nucleus) collisions, particles are 
produced in different ways. In an early stage of the collision there is a 
perturbative parton cascade process which is governed by the evolution 
equations of QCD. During a pp collision partons are produced in 
inelastic parton-parton collisions and later the partons are 
converted into hadrons with additional particle production. Here the main 
mechanism is non-perturbative: string formation and decays.

Fluctuations in the number of produced particles, which show up on the measured 
multiplicity distribution, can have several causes. One of them is certainly 
geometrical, i.e., it comes from the variation of the distribution of the 
colliding matter in space from event to event. Particle production
and especially the multiplicity distribution is very sensitive to initial
spatial distribution of matter and hence is a good place to test the baryon
junction conjecture.

From the point of view of modelling, the initial state geometry can be 
naturally implemented in the partonic version of the Glauber model 
\cite{glauber07,loi16,loi19}. Typically one chooses the initial spatial 
density of matter and, with it, computes the thickness function, 
which appears in  the expressions of the cross sections. In this 
work we will simulate the collisions in the MC-KLN event generator 
\cite{mckln}. This Monte Carlo simulator implements the $k_T$-factorization 
formalism of the Color Glass Condensate (CGC) using the KLN unintegrated gluon 
distribution \cite{kln1,kln2}. For a nucleon with a given initial spatial 
configuration, this  simulator allows one to compute  gluon production. 
Conversions to final-state 
hadrons are taken into account via the parton-hadron duality, meaning that the 
particle production at parton and hadron level only differ by a constant factor, 
also employed in previous phenomenlogical studies involving $p_T$-integrated 
observables calculated via the $k_T$-factorization formalism                        
\cite{mckln,kln1,kln2,Dumitru:2011wq,Dumitru:2012yr,Albacete:2012xq,Duraes:2016yyg}. 
With these tools we can use the baryon junction initial conditions 
and calculate the final multiplicity distribution, looking for its manifestations. 

The effect of the initial state geometry and its fluctuations on multiplicity 
distributions in pp collisions was investigated in \cite{larry16}.             
In that work the authors studied multiplicity distributions in the dense-dense 
limit of the CGC, exploring the nature of high multiplicity events in pp collisions.
They considered the standard fluctuations of the collision geometry together      
with color charge fluctuations and could obtain a reasonable description of data, 
excluding the highest multiplicity events. Additional sources of saturation scale fluctuations were required to 
explain the rare events which give rise to the tails of the multiplicity 
distribution in pp collisions.
These fluctuations are a consequence of stochastic splitting of dipoles that 
are not accounted for in the conventional frameworks of CGC \cite{Iancu:2004es,Munier:2003sj,Mueller:1993rr,Iancu:2003zr}, and are described via a log-normal distribution. The width of the 
Gaussian distribution $\sigma$, as constrained by the data, appears to be 
energy independent, suggesting that these fluctuations have a non-perturbative 
origin. In \cite{larry16} the authors did not address the specific case of the 
baryon 
junction.  More recently, the effect of BJ initial conditions in multiplicity 
distributions in pp collisions was studied in \cite{deb20}. The 
authors found that the BJ initial conditions provide a reasonable description of 
the data, but they  were not able to account for the very large multiplicity events.

In this work we use the MC-KLN simulator implementing BJ initial configurations, 
as in \cite{schenke}, with intrinsic fluctuations, as in \cite{larry16}, and apply 
it to the multiplicity distributions, as in  \cite{deb20}. Here we focus on pp and p-Pb 
collisions, the second not addressed in any of these studies, and compare our results to very recent data \cite{alice24}.

\section{Formalism}

In this section we present the ingredients of our calculation. Our strategy
is to use a well tested particle production model and vary only the initial 
conditions, covering the most accepted matter spatial distributions, and try
to determine which of them gives a better description of data. Ideally, this
procedure would rule out some initial conditions and favor others. However, 
given the complexity of the calculation, it is difficult to exclude 
configurations. Nevertheless, the results give us hints and useful insights.

\subsection{Initial conditions for the nucleon} 

\begin{figure}[h!]
\begin{tabular}{ccc}
\includegraphics[scale=0.3]{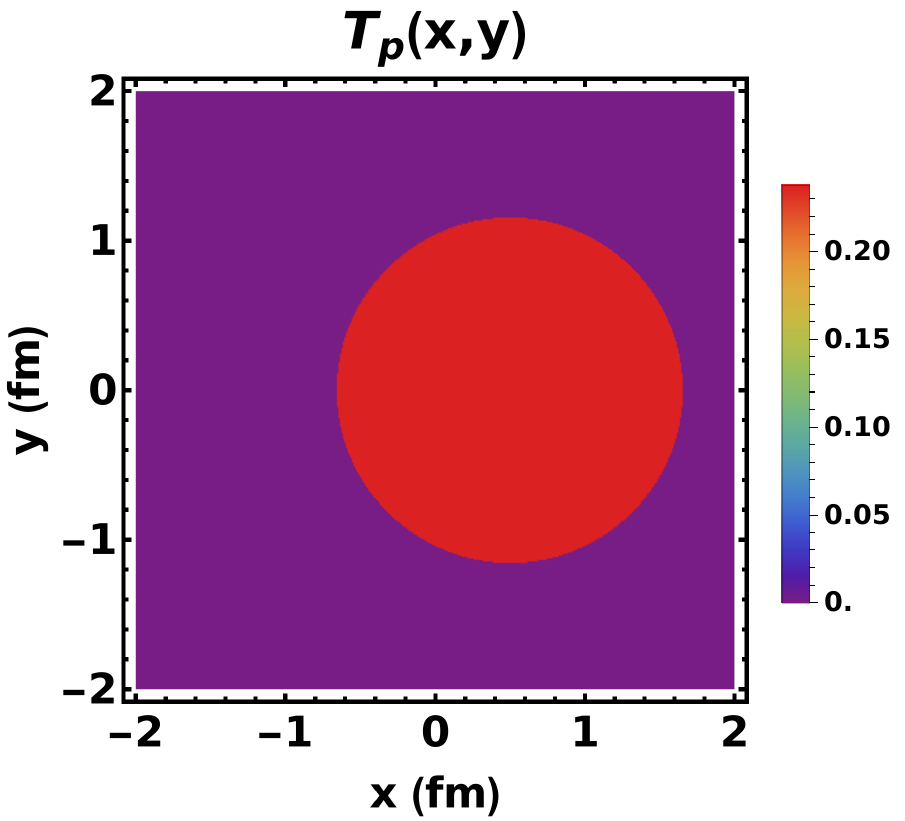}&
\includegraphics[scale=0.3]{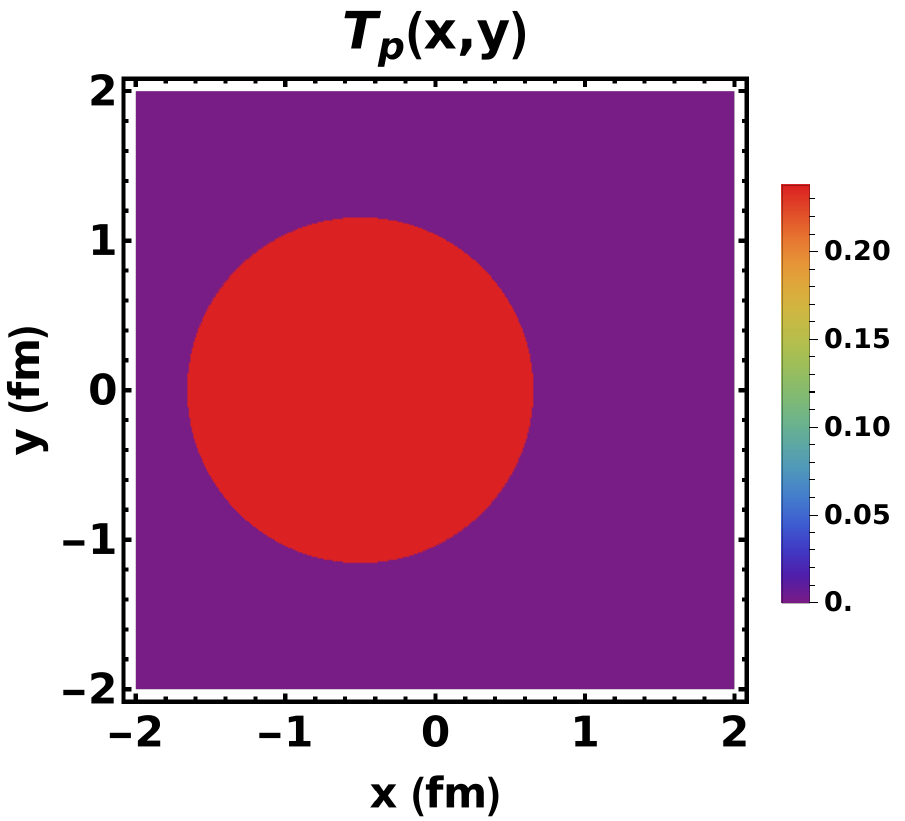} &
\includegraphics[scale=0.3]{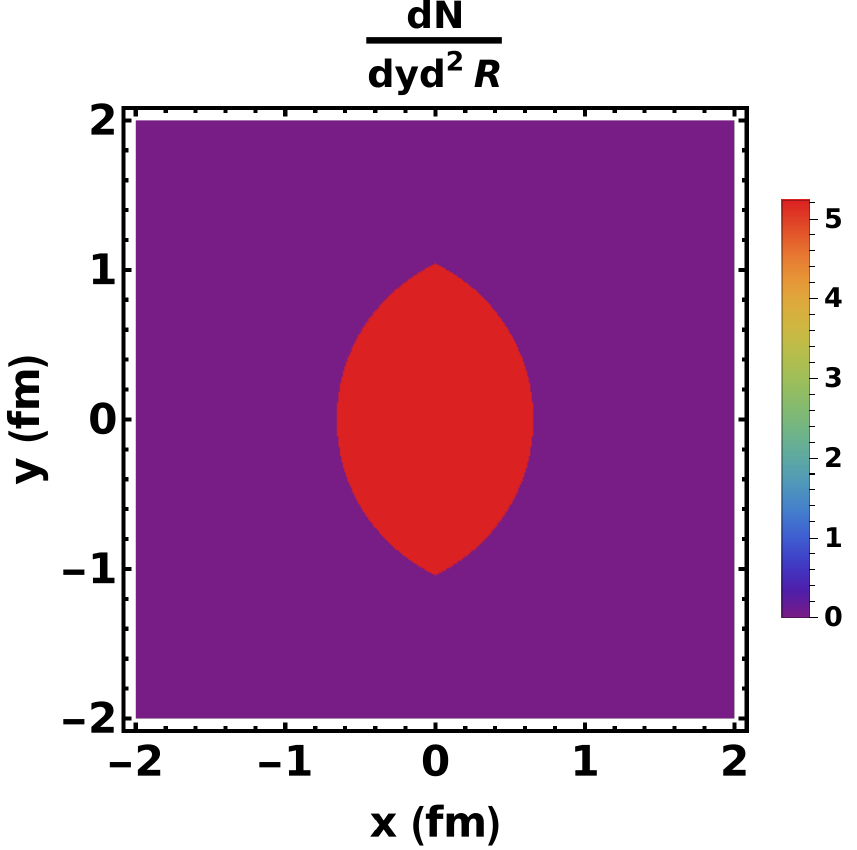}\\
\includegraphics[scale=0.3]{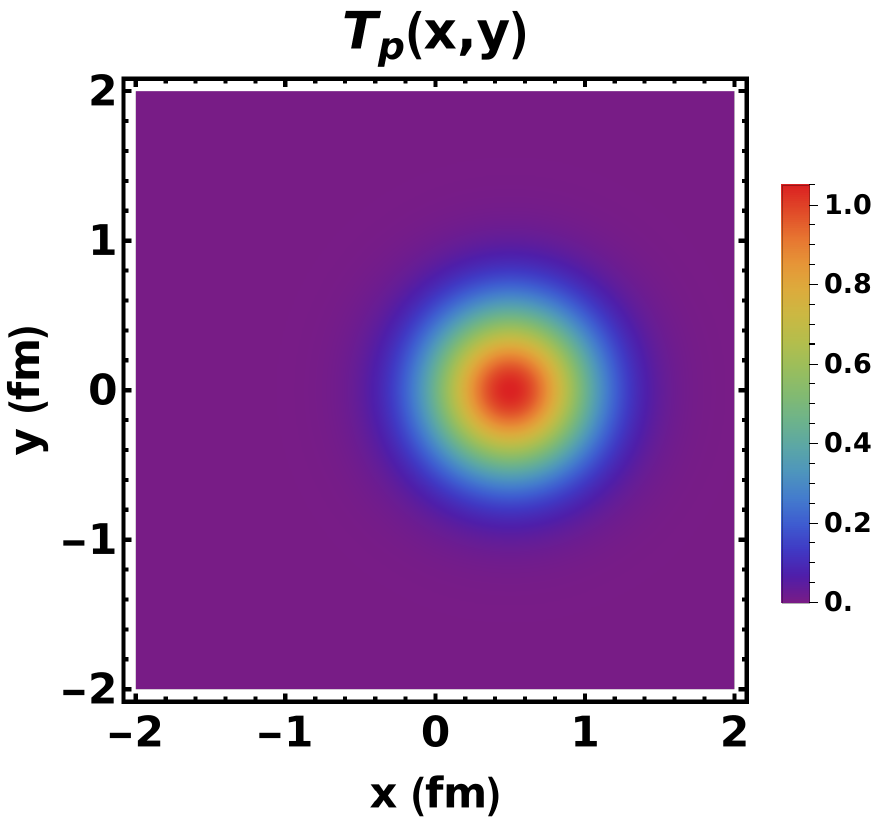}&
\includegraphics[scale=0.3]{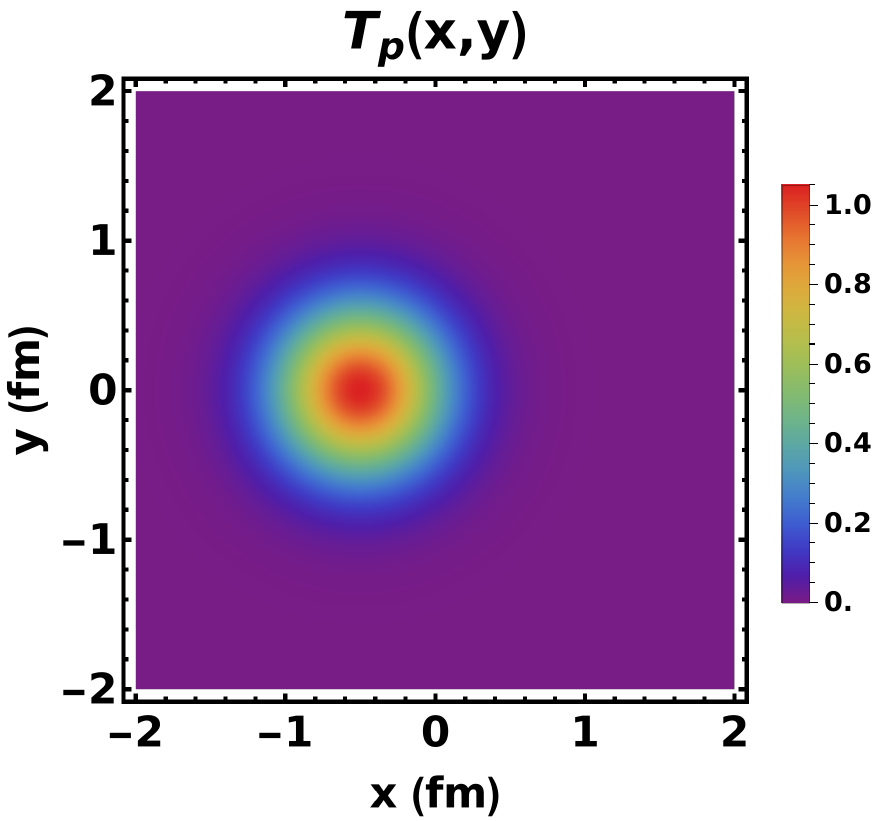}&
\includegraphics[scale=0.3]{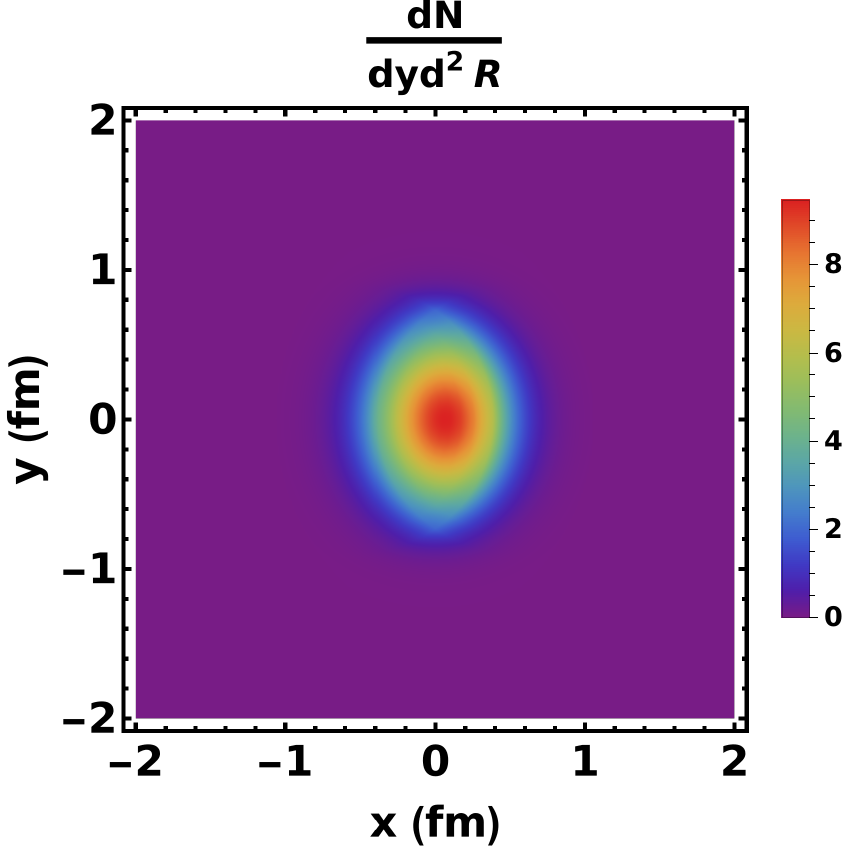}\\
\includegraphics[scale=0.3]{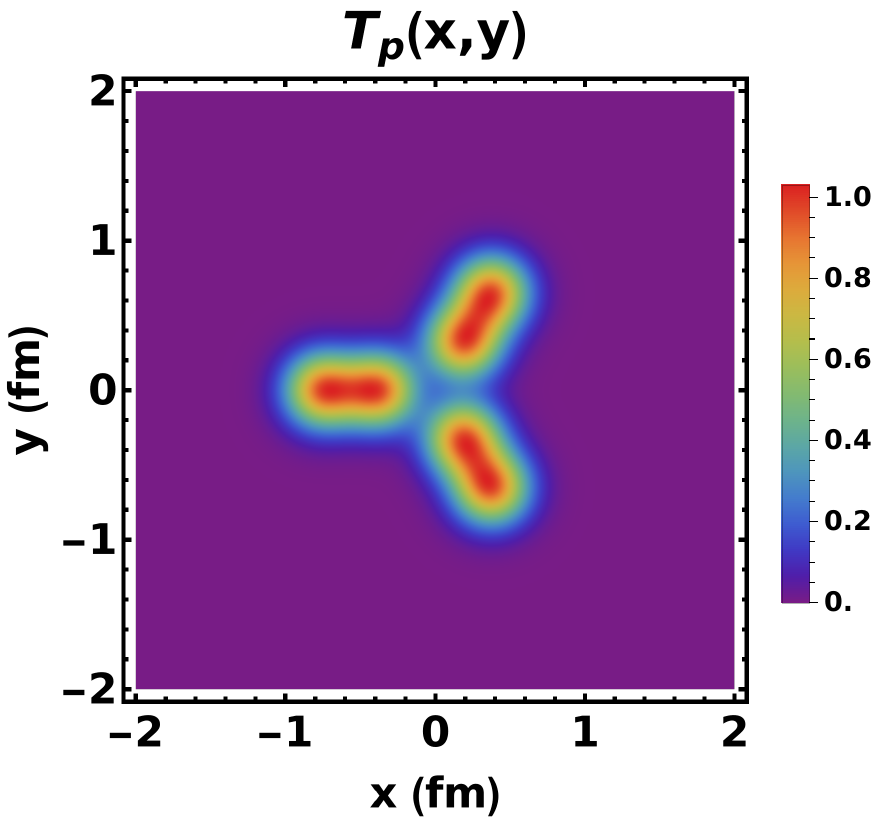}&
\includegraphics[scale=0.3]{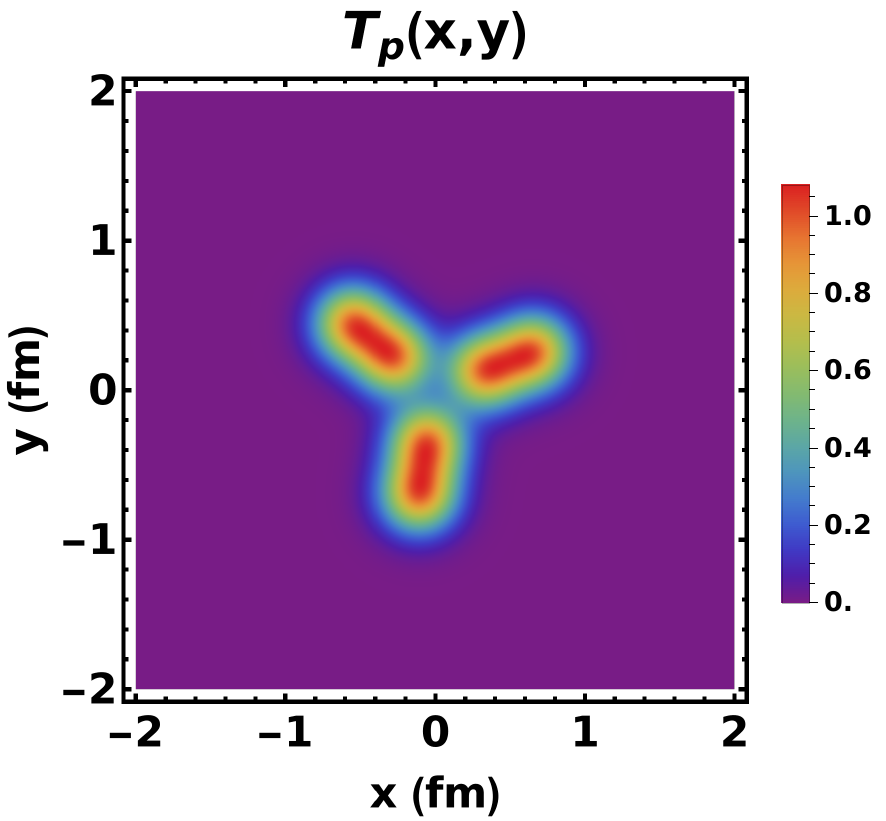} &
\includegraphics[scale=0.3]{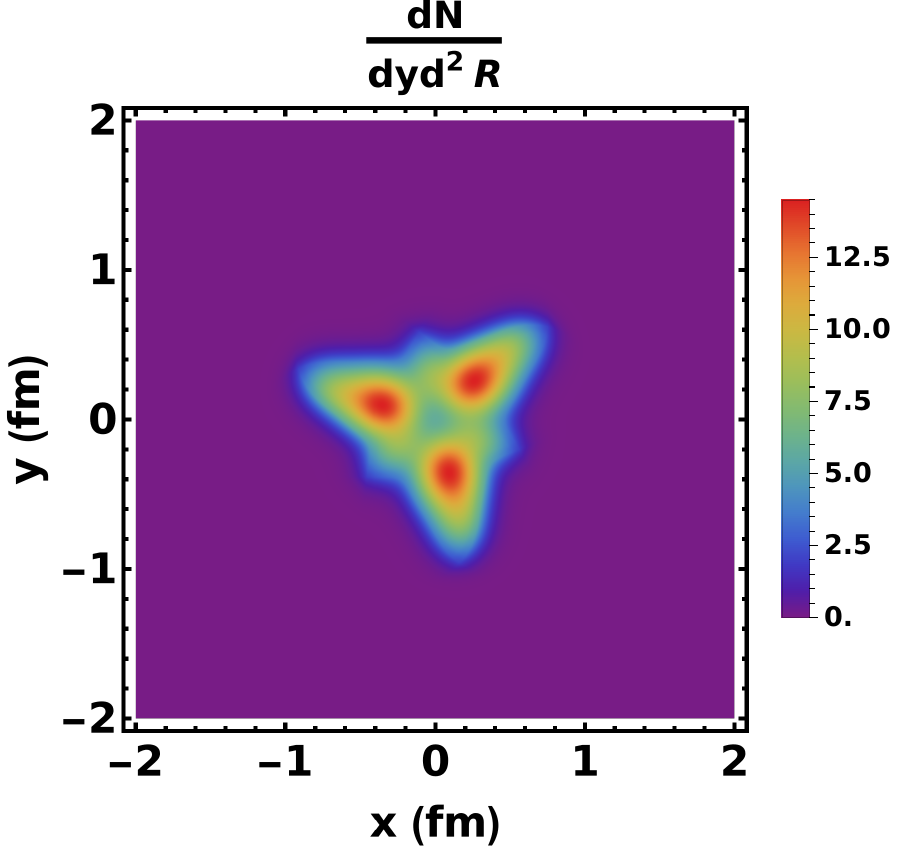}\\
\includegraphics[scale=0.3]{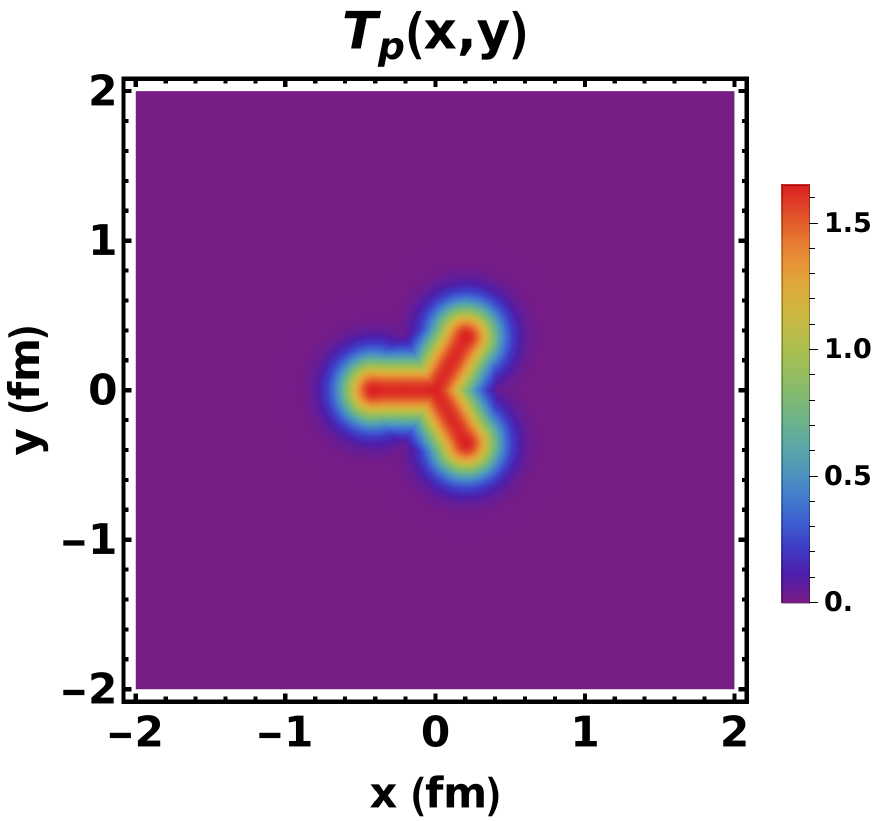}&
\includegraphics[scale=0.3]{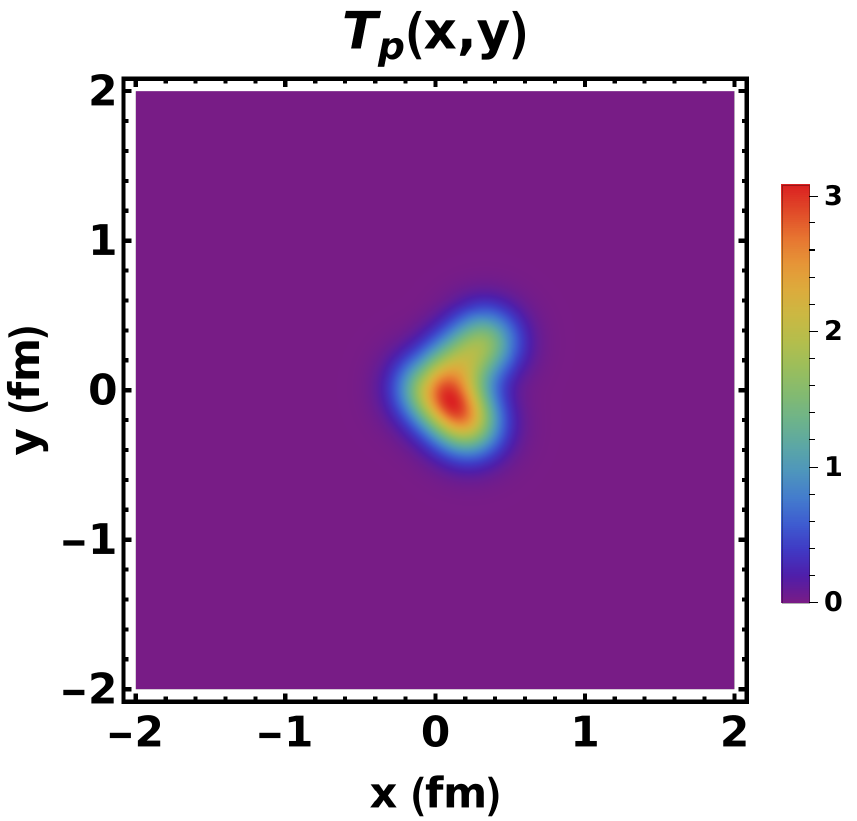} &
\includegraphics[scale=0.3]{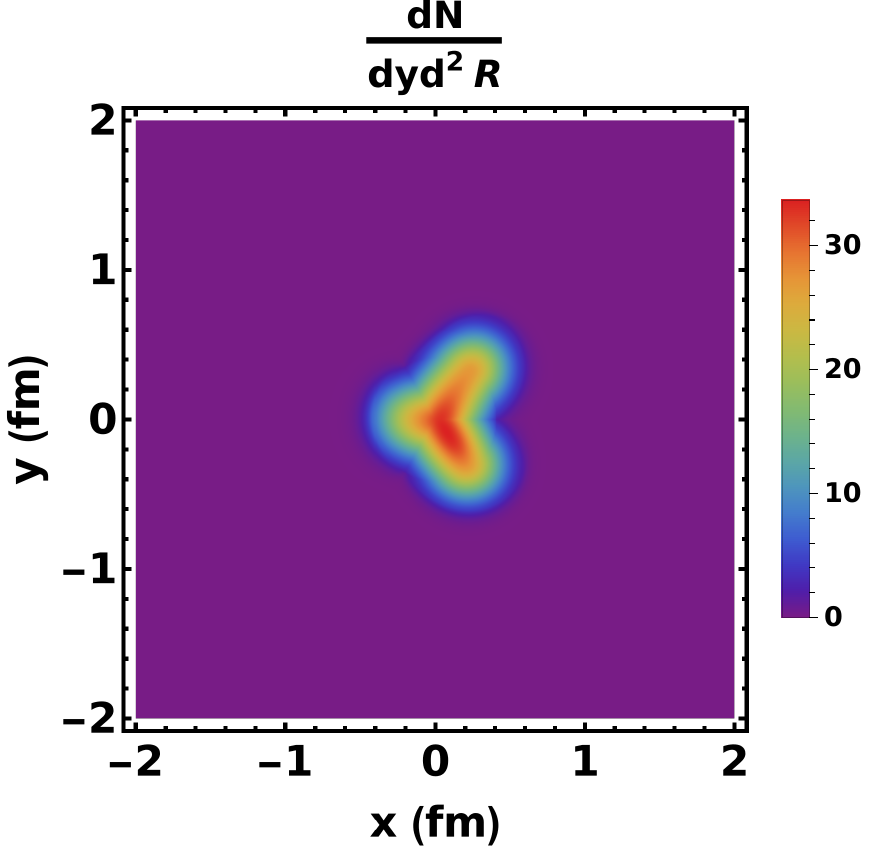}\\
\end{tabular}
\caption{Proton-proton collisions. (Left) Projectile proton thickness. 
(Middle) Target proton thickness. 
(Right) $dN/dy \, d^2 \mathbf{r}$ of the produced gluons at midrapidity. 
The rows represent different density distributions of the nucleons. From top to 
bottom: hard-sphere, Gaussian, BJ1, BJ2.}
\label{fig1}
\end{figure}

\begin{figure}[h!]
\begin{tabular}{ccc}
\includegraphics[scale=0.32]{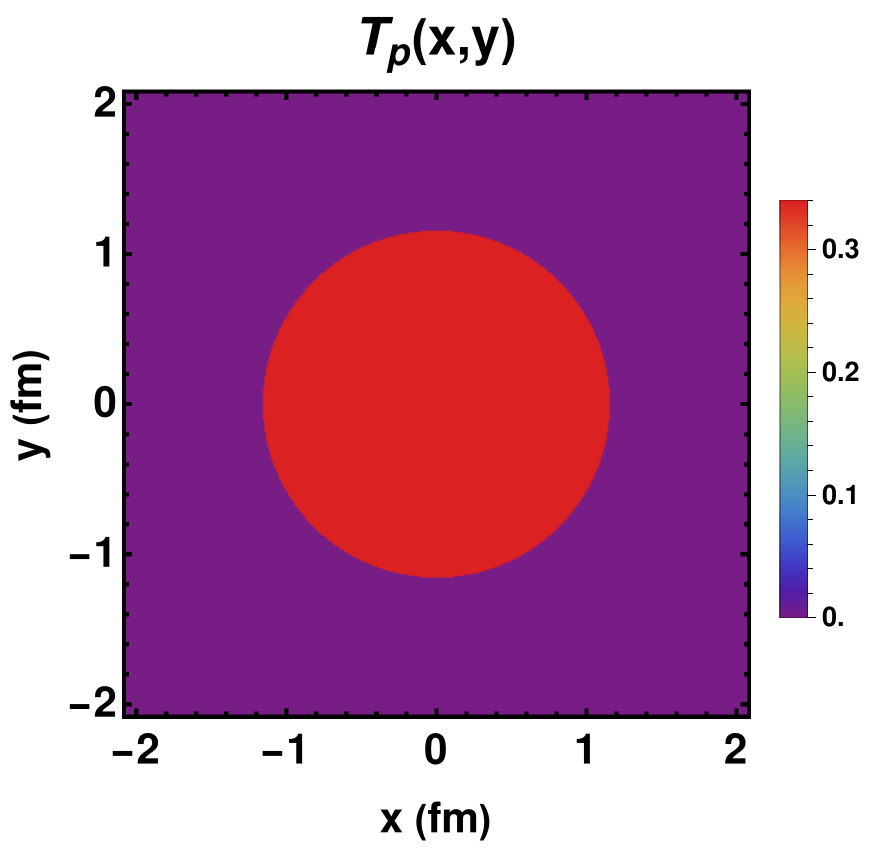}&
\includegraphics[scale=0.32]{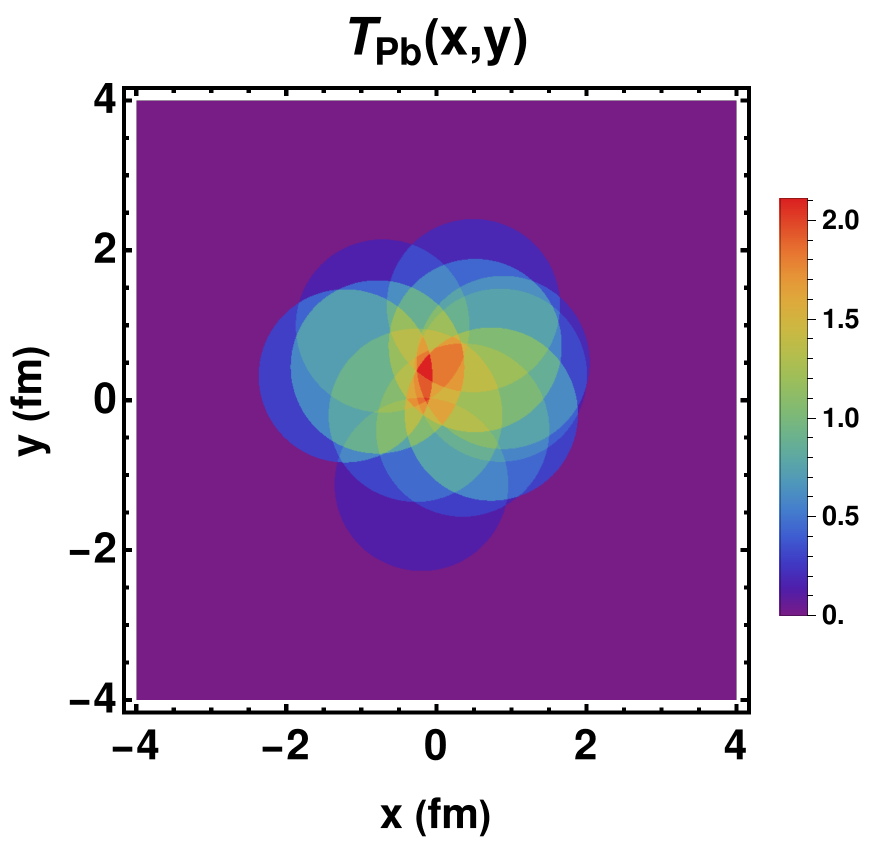} &
\includegraphics[scale=0.32]{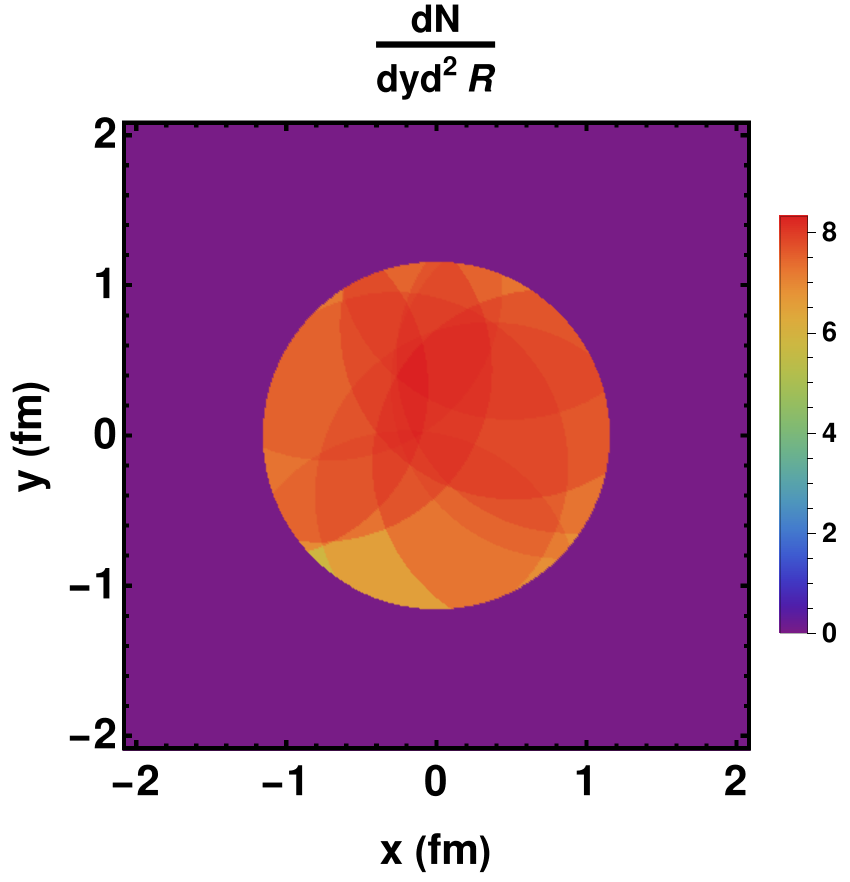}\\
\includegraphics[scale=0.32]{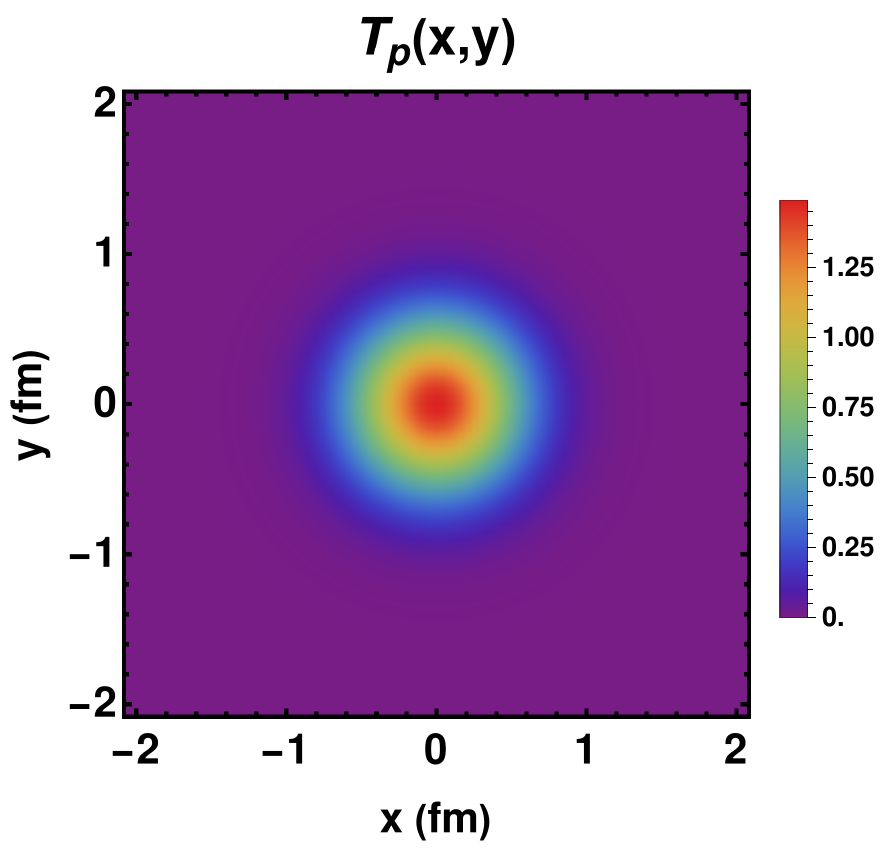}&
\includegraphics[scale=0.32]{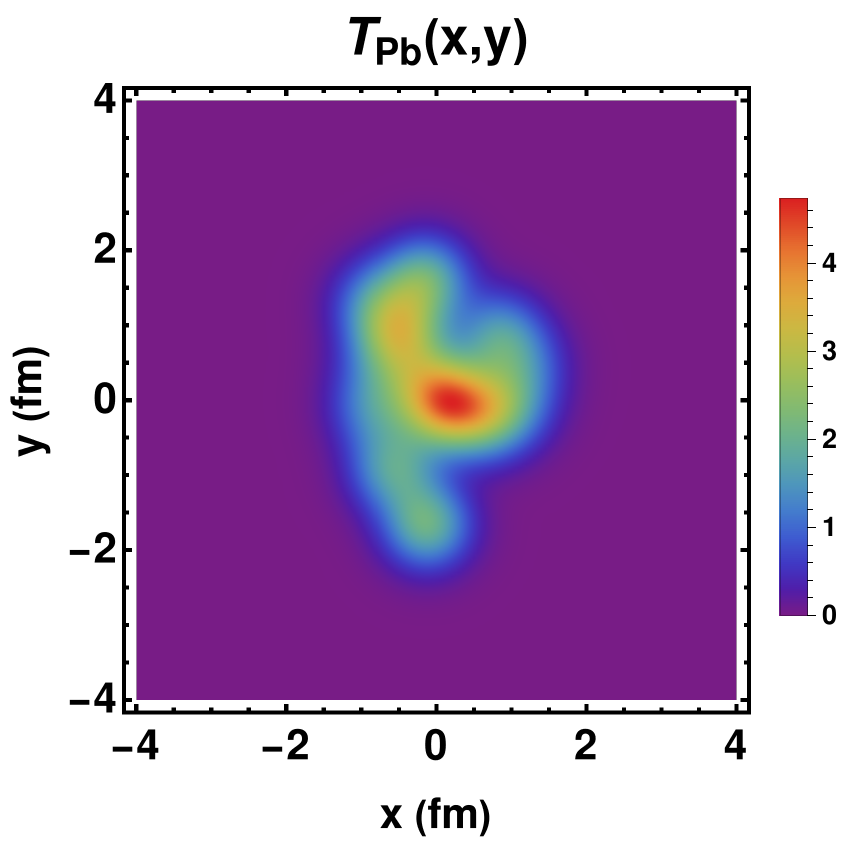}&
\includegraphics[scale=0.32]{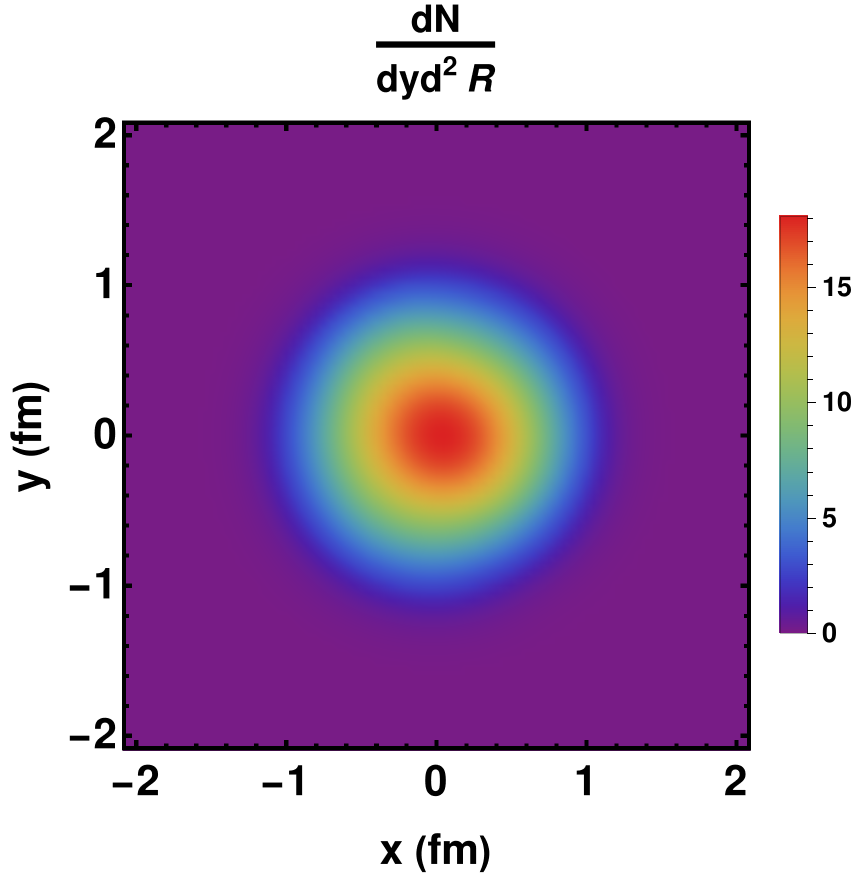}\\
\includegraphics[scale=0.32]{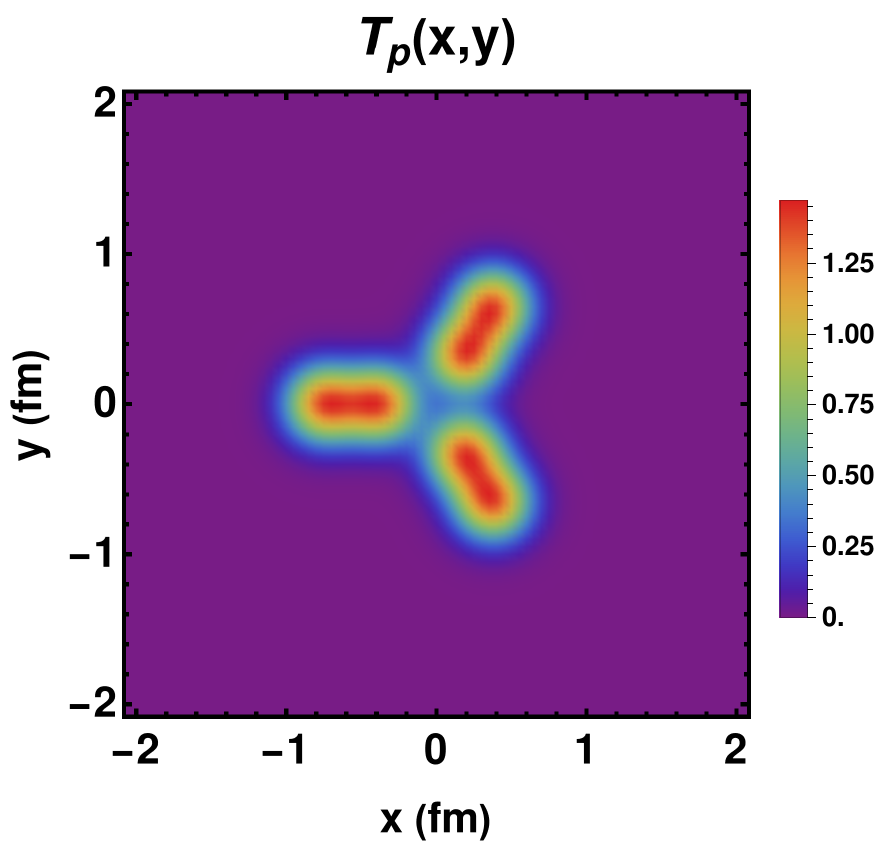}&
\includegraphics[scale=0.32]{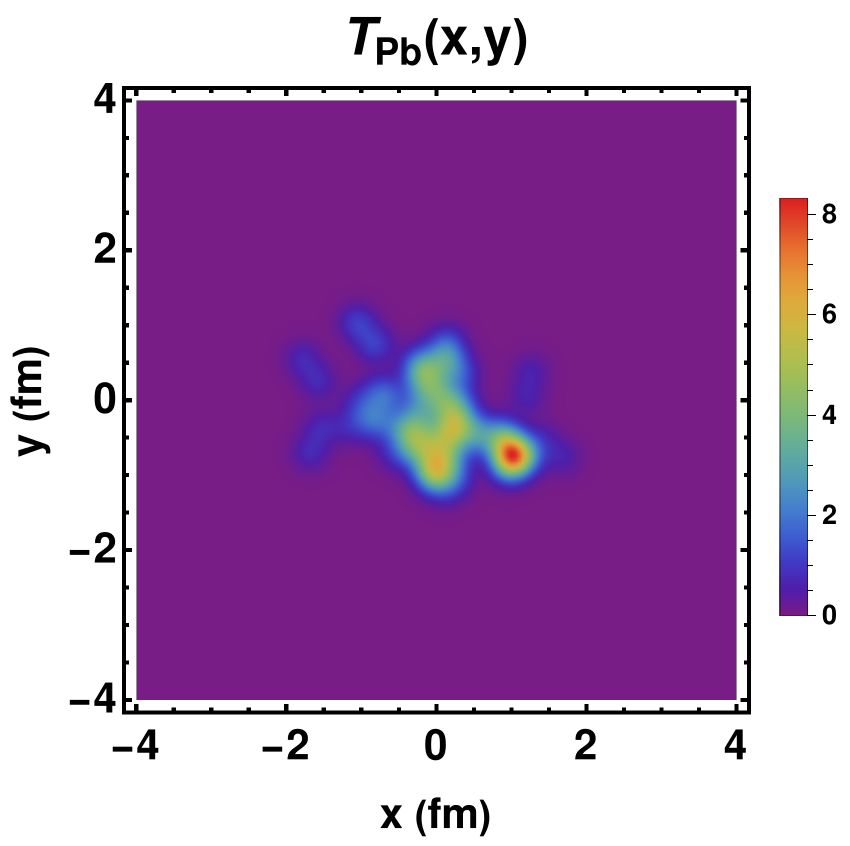} &
\includegraphics[scale=0.32]{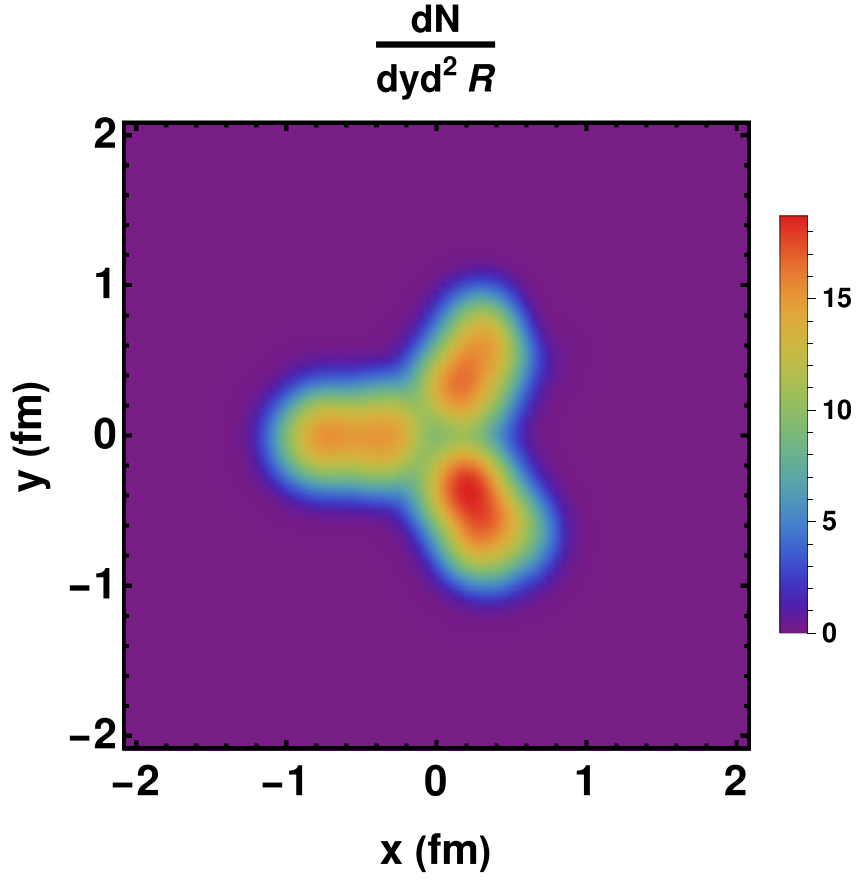}\\
\includegraphics[scale=0.32]{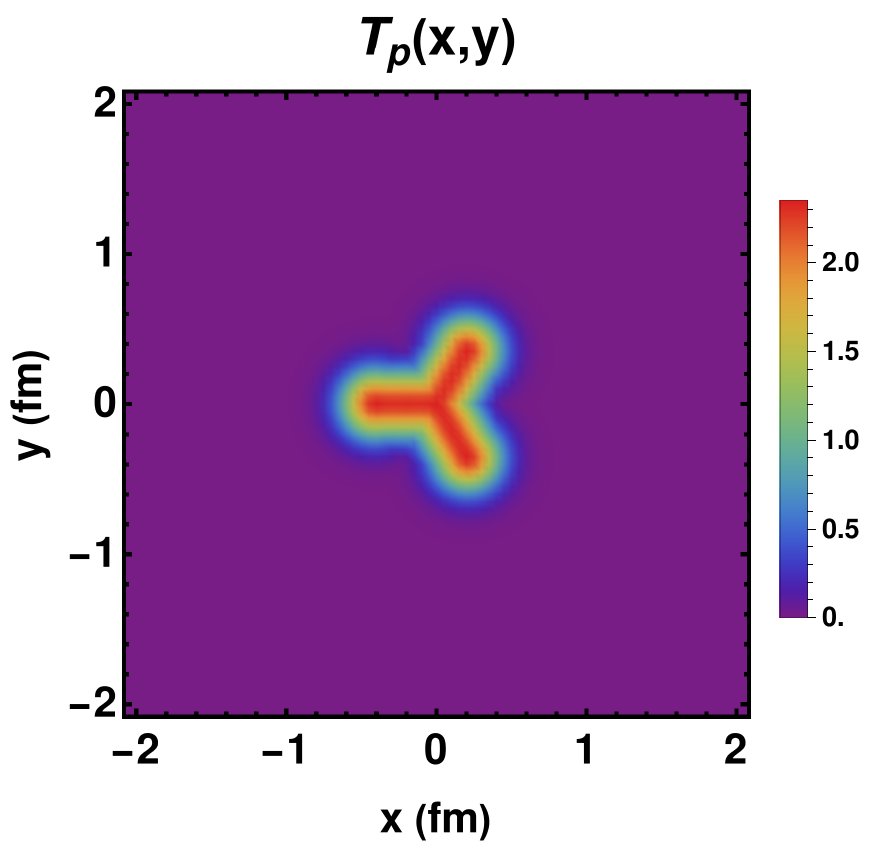}&
\includegraphics[scale=0.32]{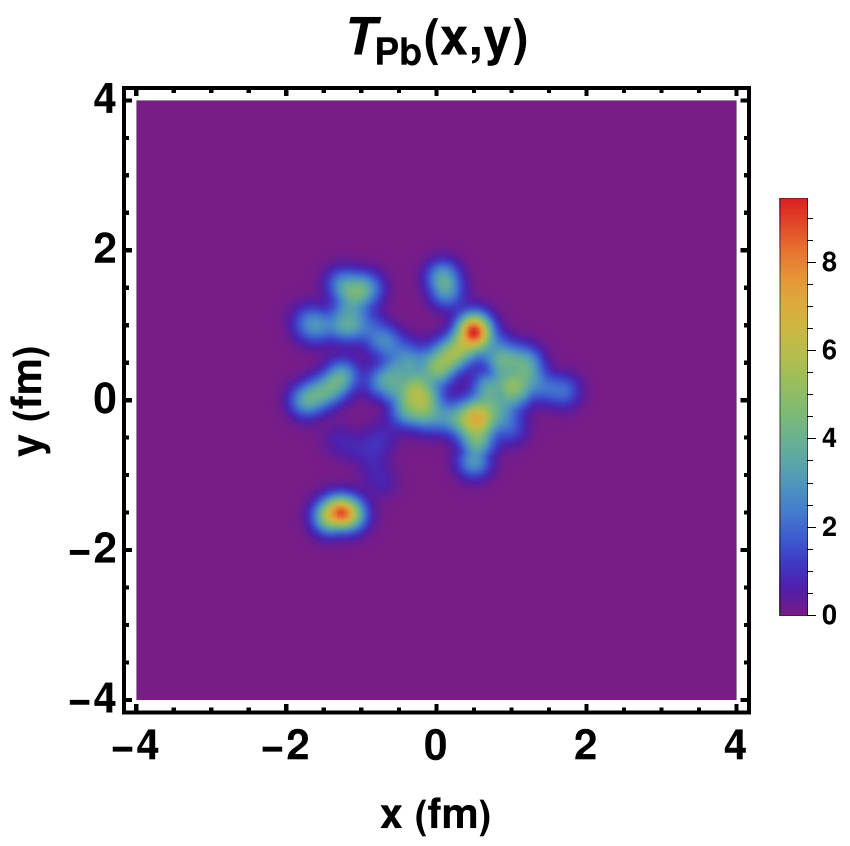} &
\includegraphics[scale=0.32]{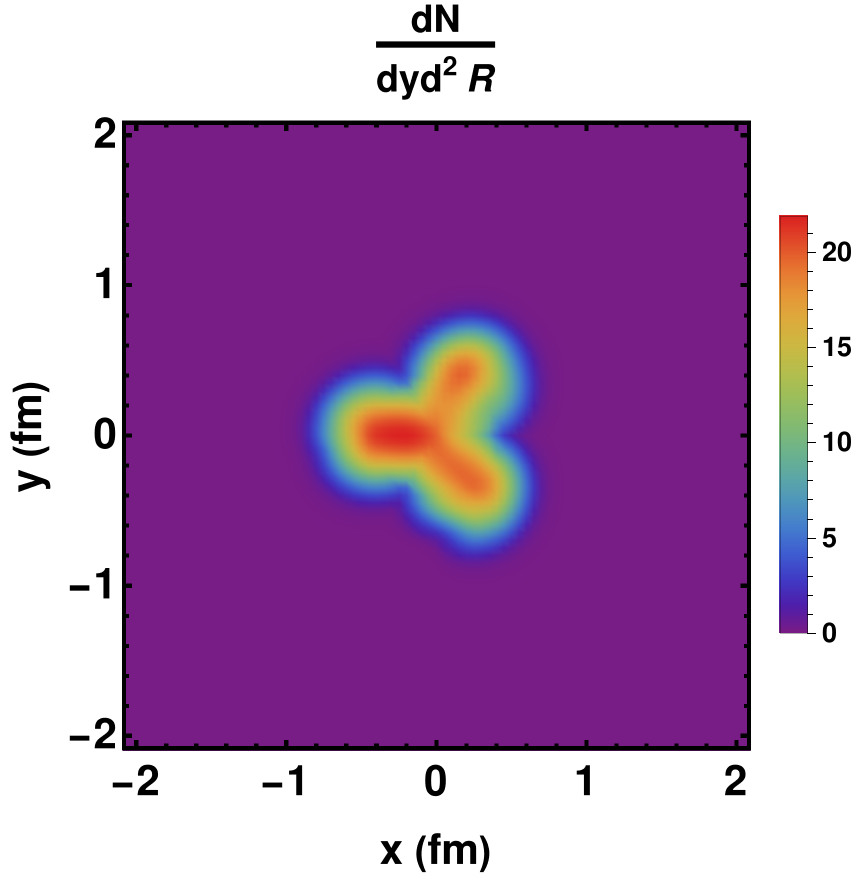}\\
\end{tabular}
\caption{Proton-lead collisions. (Left) Proton thickness. 
(Middle) Lead participant thickness. 
(Right) $dN/dy \, d^2 \mathbf{r}$ of the produced gluons at midrapidity. 
The rows represent different density distributions of the nucleons. From top to 
bottom: hard-sphere, Gaussian, BJ1, BJ2.}
\label{fig2}
\end{figure}

We start by introducing  four types of nucleon matter spatial distributions 
used in our simulations: hard-sphere, Gaussian, analytical baryon junction
(from now on called BJ1) and a slightly different and purely numerical baryon 
junction (from now on called BJ2). Examples of the  thickness function 
(z-integrated density) of each configuration are shown on the left panels of 
Figs. \ref{fig1} and \ref{fig2}. The configurations are described below.

\subsubsection{Hard-sphere}

The hard-sphere is based on the collision criterion of distances between 
two nucleon centers \cite{Albacete:2012xq}. 
Explicitly, there is a collision when the  
distance between the projectile (proj) and the target (targ) is:
\begin{equation}
d^2= \big(x_{proj}-x_{targ} \big)^2 + \big(y_{proj}-y_{targ} \big)^2 
\le\frac{\sigma_0}{\pi} \, ,
\end{equation}
where $\sigma_0$ is the transverse area of the large $x$ partons of a nucleon.
As in Ref. \cite{Albacete:2012xq}, we choose the parameter $\sigma_{0}$ to be equal 
to $42$ mb. This collision criterion implies that each 
nucleon is geometrically described by a filled disk, with thickness given by:

\begin{equation}
	T(\mathbf{r}_\perp)=  \frac{\Theta(\sqrt{\sigma_0/\pi} 
- r_\perp)}{\sigma_0} \, ,
\end{equation}
where $\Theta(x)$ is the usual Heaviside step function.

\subsubsection{Gaussian nucleon}

A way to soften the sharp edges of the hard-sphere model while still not    
including sub-nucleonic degrees of freedom and also maintaining the overall 
smooth shape of the proton is to consider a Gaussian nucleon. In this case, 
the nucleon  transverse profile can be calculated analytically and is given 
by

\begin{equation}
    T_N(\mathbf{r}_\perp)=\frac{e^{-\frac{\mathbf{r}_\perp^2}{2B_p}}}{2\pi B_p}\, ,
    \label{thicknessgauss}
\end{equation}
with $B_p$ being the nucleon width, which is related to the root mean square    
radius of the proton as $r_p=\sqrt{2B_p}$. Older fits to HERA data~\cite{reza},  
as well as a recent Bayesian analysis based solely on information from $e+p$ 
collisions from HERA~\cite{Mantysaari:2022ffw}, show that $                    
B_p\approx 4\,{\rm GeV}^{-2}$, while a recent Bayesian analysis of combined data 
on diffractive $J/\Psi$ photoproduction in $\gamma + p$ and $\gamma + Pb$ collisions 
yields a smaller value~\cite{Mantysaari:2025ltq}.

The probability of having a collision at nucleon level between a projectile and a  
target separated by an impact parameter $b$ is now related to their overlap 
function, $T_{NN^\prime}$, which can be calculated analytically in this model,
\begin{equation} 
P(b)=1-\exp\big( -\sigma^{eff} T_{NN^\prime}(b)\big)\,,\qquad T_{NN^\prime}(b) = 
\int T_N(x-b/2,y) T_{N^\prime}(x+b/2,y) dxdy = \frac{e^{-b^2/4B_p}}{4 \pi B_p}\,,
\label{overlapgauss}
\end{equation}
where $\sigma_{eff}$ is fixed so that the integral of $P(b)$ over the impact    
parameter yields the inelastic nucleon-nucleon cross-section at a given energy. 
The thickness and overlap functions from Eqs. (\ref{thicknessgauss}) and  
(\ref{overlapgauss}) enter as inputs in the MC-KLN.

\subsubsection{Baryon junction BJ1}

The BJ1 is an analytical Ansatz for the characteristic ``Y'' shape of the 
proton. 
The  three effective quarks are disposed in the vertices of an 
equilateral triangle and they have coordinates $\mathbf{r}_1$,  
$\mathbf{r}_2$ 
and $\mathbf{r}_3$. To these quarks we add three gluon distributions with  
peaks at
half of the distances between the nucleon center and the quark peaks. The 
resulting thickness is explicitly given by:

\begin{equation}
	T_N (x,y)=T_N^q (x,y)+ T_N^g (x,y) \, ,
	\label{Tn}
\end{equation}
with the quark and gluon hot-spots, respectively, given by:
\begin{equation}
	T_N^q(x,y)= \frac{(1-\kappa)}{6\pi B_h^2} \sum_{i=1}^3 
\exp \Bigg(-\frac{\big(x-(x_p+x_i)\big)^2+\big(y-(y_p+y_i)\big)^2}{2B_h^2}\Bigg)\, ,
	\label{Tnq}
\end{equation}
and
\begin{equation}
	T_N^g(x,y)=\frac{\kappa}{6\pi B_h^2} \sum_{i=1}^3  \exp              
\Bigg(-\frac{\big(x-(x_p+\frac{x_i}{2})\big)^2 + 
\big(y-(y_p+\frac{y_i}{2})\big)^2}{2B_h^2}\Bigg)\, ,
	\label{Tng}
\end{equation}
In the formulas above, $\mathbf{r}_p=(x_p,y_p)$ is the coordinate of the 
center of the nucleon in the transverse plane, $\kappa=0.5$ is 
the gluon fraction and $B_h=0.17 \text{ fm}$ is the hot-spot size. All these 
parameters were fixed in previous works 
\cite{nosso,deb20,schenke,disserta,glazek-p}. 

Each quark is positioned according to 
\begin{equation}
	\mathbf{r}_i=\frac{d}{2}\bigg(\cos(\phi_i + \alpha)\sin(\theta)\text{,} 
\sin(\phi_i + \alpha)\sin(\theta) \text{,} \cos(\theta) \bigg) \, ,
\label{ris}
\end{equation}
with $\phi_1 = \pi/3$, $\phi_2 = -\pi/3$, $\phi_3 = - \pi$ and $d=1.5$ fm.  
The $\alpha$ parameter is randomly chosen in the range $[0,2\pi]$, giving  
rotational freedom to the entire structure in the transverse plane, 
and the azimuthal $\theta$ is randomly chosen in the range $[0,\pi]$.
The respective overlap is calculated as:
\begin{equation}
T_{NN^\prime}(b) = \int T_N(x-b/2,y) \,  T_{N^\prime}(x+b/2,y) dx\, dy 
 = T_{NN^\prime}^{qq}(b)\, + \, T_{NN^\prime}^{gg}(b)+
T_{NN^\prime}^{gq}(b)\, + \, T_{NN^\prime}^{qg}(b),
\label{Tnn}
\end{equation}
and hence it can be split into quark-quark, gluon-gluon, gluon-quark and 
quark-gluon overlaps. Explicitly, each term is given by:

\begin{eqnarray}
T_{NN^\prime}^{qq}(b) &=& \int T_N^q(x-b/2,y) T_{N^\prime}^q(x+b/2,y) dxdy
 \nonumber \\
                     &=& \frac{ (1-\kappa)^2}{36\pi B_h^2}          
\sum_{i=1}^3\sum_{j=1}^3 \exp \bigg(-\frac{\big(b+(x_p+x_i)-(x_a+x_j)\big)^2
+\big((y_p+y_i)-(y_a+y_j)\big)^2}{4 B_h^2} \bigg)\,,
\label{Tqq}\\
T_{NN^\prime}^{gg}(b) &=& \int T_N^g(x-b/2,y) T_{N^\prime}^g(x+b/2,y) dxdy 
\nonumber\\
                     &=& \frac{ \kappa^2 }{36\pi B_h^2} \sum_{i=1}^3\sum_{j=1}^3  
\exp \Bigg(- \frac{\big(b+(x_p+x_i/2)-(x_a+x_j/2)\big)^2+ \big((y_p+y_i/2) - 
(y_a+y_j/2)\big)^2}{4B_h^2}   \Bigg)\,,
\label{Tgg}\\
T_{NN^\prime}^{gq}(b) &=& \int T_N^g(x-b/2,y) T_{N\prime}^q(x+b/2,y) dxdy\nonumber
 \\
&=& \frac{\kappa (1-\kappa)}{36\pi B_h^2} 
\sum_{i=1}^3 \sum_{j=1}^3 \exp \bigg(-\frac{\big(b+ (x_p+x_i/2) -(x_a+x_j) \big)^2 
+ \big( (y_p+y_i/2)-(y_a+y_j)\big)^2}{8B_h^2}\bigg)\,,
\label{Tgq}\\
T_{NN^\prime}^{qg}(b) &=& \int T_N^q(x-b/2,y) T_{N^\prime}^g(x+b/2,y) dxdy 
\nonumber \\
&=& \frac{\kappa (1-\kappa)}{36\pi B_h^2} 
\sum_{i=1}^3 \sum_{j=1}^3 \exp \bigg(-\frac{\big(b+ (x_p+x_i) -(x_a+x_j/2) \big)^2 
+ \big( (y_p+y_i)-(y_a+y_j/2)\big)^2}{8B_h^2}\bigg)\,.
\label{Tqg}
\end{eqnarray}
In the formulas above we introduced the position of the center of the target  
in the transverse plane, $\mathbf{r}_a=(x_a,y_a)$, and the peaks of the quark 
distribution in the target nucleon, $j=1,2,3$, as in  Eq.(\ref{ris}). 
We use the analytical thickness (Eq.(\ref{Tn})) and 
overlap (Eq.(\ref{Tnn})) as inputs in the MC-KLN.

\subsubsection{Baryon Junction BJ2}

The baryon junction BJ2 is a numerical Ansatz for the Y-like structure, as 
employed in \cite{schenke}. In this approach we draw the position of the three 
quarks in the 3D space according to the distribution:
\begin{equation}
P(\mathbf{r})=\frac{e^{-\frac{\mathbf{r}^2}{2B_r}}}{\big(2\pi B_r\big)^{3/2}} \,,
\end{equation}
with $B_r=0.41^2 \text{ fm}^2$ as the characteristic size of the proton.   
After that we calculate the central point (Fermat point) as the point that 
simultaneously minimizes the distance to the three quarks. In the last step 
the gluon flux tubes are filled by Gaussians with characteristic squared
radius $B_t=0.15^2 \text{ fm}^2$ and the distribution is normalized to 1.   
The parameters were taken from \cite{schenke} and the overlap of the string 
junction will be approximated by the overlap of the Gaussian nucleon of 
Eq.(\ref{overlapgauss}), but with constant $B_r$.

\subsection{Initial condition for the nucleus}

The initial condition for the lead nucleus will be defined  by randomly 
positioning 208 nucleons inside the nucleus according to the Woods-Saxon profile 
\begin{equation}
    \rho(r)=\rho_0\frac{1+w(r/R)^2}{1+\exp\big(\frac{r-R}{a}\big)} \, ,
    \label{nuclear charge density}
\end{equation}
where $\rho_0$ is the nucleon density in the center of the nucleus, $R$ is the 
radius of the nucleus, $w$ represents deviations from the sphere format, and 
$a$ is the skin depth. Some usual values for these parameters can be found in 
\cite{Pbparam}. We perform our simulation by sorting 208 nucleon center 
positions,           
$\mathbf{r}_n$, and then inserting the chosen nucleon distribution around it. 
In the end, the Pb thickness is:
\begin{equation}
	T_{Pb} (x,y)=\sum_{n=1}^{208} T_n (x,y)  \, ,
	\label{TPb}
\end{equation} 
with $T_n$ as the thickness of the chosen type of nucleon. Examples of the 
(participant) thickness for the lead nucleus are shown in the second column of Fig. 
\ref{fig2}.

\subsection{$k_T$-factorization}

In order to compute the rapidity and transverse momentum distributions of the 
particles produced in the final state, we use the $k_T$-factorization. 
For details we refer the reader to Ref.  \cite{paperrcbk,kt2011,kt2008}.

The single-inclusive (small-x) gluon production cross-section is given by:
\begin{eqnarray}
\frac{d\sigma^{A+B \rightarrow g}}{dy \, d^2p_T \, d^2 \mathbf{r_{\perp}}} = 
K \, \frac{2}{C_F\text{ } p_T^2}
\int^{p_T} \frac{d^2k_T}{4} \int d^2b \text{ } \alpha_s(Q) \, 
\phi_p \bigg(\frac{|p_T+k_T|}{2},x_1;b \bigg) \, \phi_T 
\bigg(\frac{|p_T-k_T|}{2},x_2;R-b \bigg) \, .
    \label{ktfactorization}
\end{eqnarray}
In our notation $K$ is a multiplicative constant introduced to account for  
higher-order corrections. In practical terms, it will be fixed by requiring      
that the mean number of charged particles is reproduced. $p_T$ is the transverse 
momentum of the produced gluon, 
$k_T$ is the intrinsic transverse momentum and $C_F=(N_c^2-1)/2N_c$ is the 
color factor. The $2\rightarrow1$ kinematics implies that:
\begin{equation}
x_{1,2}=\frac{p_T}{\sqrt{s}}e^{\pm y}  \, .
\end{equation}
The unintegrated gluon distribution function (UGD) gives the number of gluons 
per $x$, per transverse momentum unit. In this work we use the KLN UGD, which is 
given by:
\begin{equation}
\phi_{KLN}(\mathbf{k},x) = \frac{2C_F}{3\pi^2}\frac{(1-x)^4}{\alpha_s}  
\left \{ \begin{matrix} 1, & \text{, if } k\le Q_s \\ 
\big(Q_s/k\big)^2 & \text{, if } k> Q_s \end{matrix} \right. 
    \label{KLN}
\end{equation}
The UGD is evaluated  at a  saturation scale given by \cite{Q2KLN}

\begin{equation}
Q_s^2(x,\mathbf{r_{\perp}})=T(\mathbf{r_{\perp}})\,\,               
\frac{2\text{ GeV}^{ 2}}{1.53 \text{ fm}^{-2}}\,\, 
\bigg(\frac{0.01}{x} \bigg)^{\bar{\lambda}} \, ,
    \label{Q2}
\end{equation}
where $T(\mathbf{r_{\perp}})$ is the projectile or target thickness and 
$\bar{\lambda}=0.23$.

The MC-KLN  convolutes the initial conditions of p and p (Pb) with the       
$k_T$-factorization formula of Eq.(\ref{ktfactorization}) by calculating       
$dN/dy \,d^2 \mathbf{r_{\perp}}$ of gluons at a given rapidity in the overlap 
region. One example of each initial state geometry for pPb collisions is shown in 
Fig.\ref{fig2}, for the same seed. On the left column we present the proton      
thickness, on the middle column we show the thickness of the lead nucleus and 
on the right column we show the convolution of the initial conditions and
Eq.(\ref{ktfactorization}) at midrapidity. 

In Fig.\ref{fig2} we can observe that, remarkably,    
the produced gluon distribution on the right column inherits the proton shape, 
which makes pPb collisions good systems to probe the proton initial geometry. 
This happens  because the proton size, which is about $1\text{ fm}$, is much    
smaller than the lead radius ($6.62 \text{ fm}$). Thus, the overlap region will 
invariably present the proton shape. On the other hand, in pp collisions, since 
the projectile and    
target are comparable in size, only fragments of the proton structure will overlap. 
More specifically, as represented in Fig.\ref{fig1}, for hard-sphere and Gaussian  
initial conditions the convolution will only inherit the proton structure for $b=0$ 
events, and for the baryon junctions, only for specific angle combinations. Since 
these conditions are rare, in general this system become less sensitive to the 
initial proton geometry than pPb.

Finally, in order to calculate the final charged particle distribution, we 
assume  parton-hadron duality. This means that the number of detected   
hadrons is assumed to be proportional to the number of produced gluons. 
We also use the  $y\rightarrow\eta$ jacobian
\begin{equation}
\frac{\cosh \eta}{\sqrt{\cosh^2\eta+m_\pi^2/k^2}} 
\end{equation}
to obtain the pseudorapidity distribution. In the above expression 
$m_\pi\approx 140 \text{ MeV}$ is the pion mass. We integrate the distribution  
(\ref{ktfactorization}) in $d^2 p_T $ and in the transverse plane to obtain the 
particle production at a given (pseudo-)rapidity value.

\subsection{Intrinsic fluctuations}

So far we have considered only geometrical fluctuations, i.e., those arising 
from changing positions of the initial quarks and gluons in the transverse 
plane. In order to describe  high-multiplicity events, we must include     
dynamical fluctuations in the saturation scale \cite{larry16} of a given 
nucleon in the collision. These fluctuations are characterized by deviations 
of the saturation scale from its mean value and they follow the distribution:
\begin{equation}
P(\ln(Q^2 / \left \langle Q^2 \right \rangle)) = \frac{1}{\sqrt{2\pi} 
\sigma} \exp \Bigg(-\frac{\ln^2(Q(\mathbf{s}_\perp)^2 / \left \langle 
Q(\mathbf{s}_\perp)^2 \right \rangle)}{2\sigma^2} \Bigg) \, ,
    \label{dinamica}
\end{equation}
with $\sigma$ as a free parameter to be adjusted. These fluctuations are such 
that large deviations from the averaged value are rare, which implies that 
high-multiplicity events are rare too.

\section{Results}

We will now address the recent ALICE data\cite{alice24} on 
charged particle multiplicity distributions measured in pp collisions at 
$\sqrt{s}=2.76\text{, }5.02\text{, }7\text{, }8 \text{, } 13\text{ TeV} $ 
and pPb collisions at 
$\sqrt{s}=5.02\text{, }8.16 \text{ TeV} $, in the range 
$0.15<p_T<10 \text{ GeV/c}$ and $-0.8 < \eta < 0.8$.

\begin{figure}[h!]
\begin{tabular}{cc}
\includegraphics[scale=0.28]{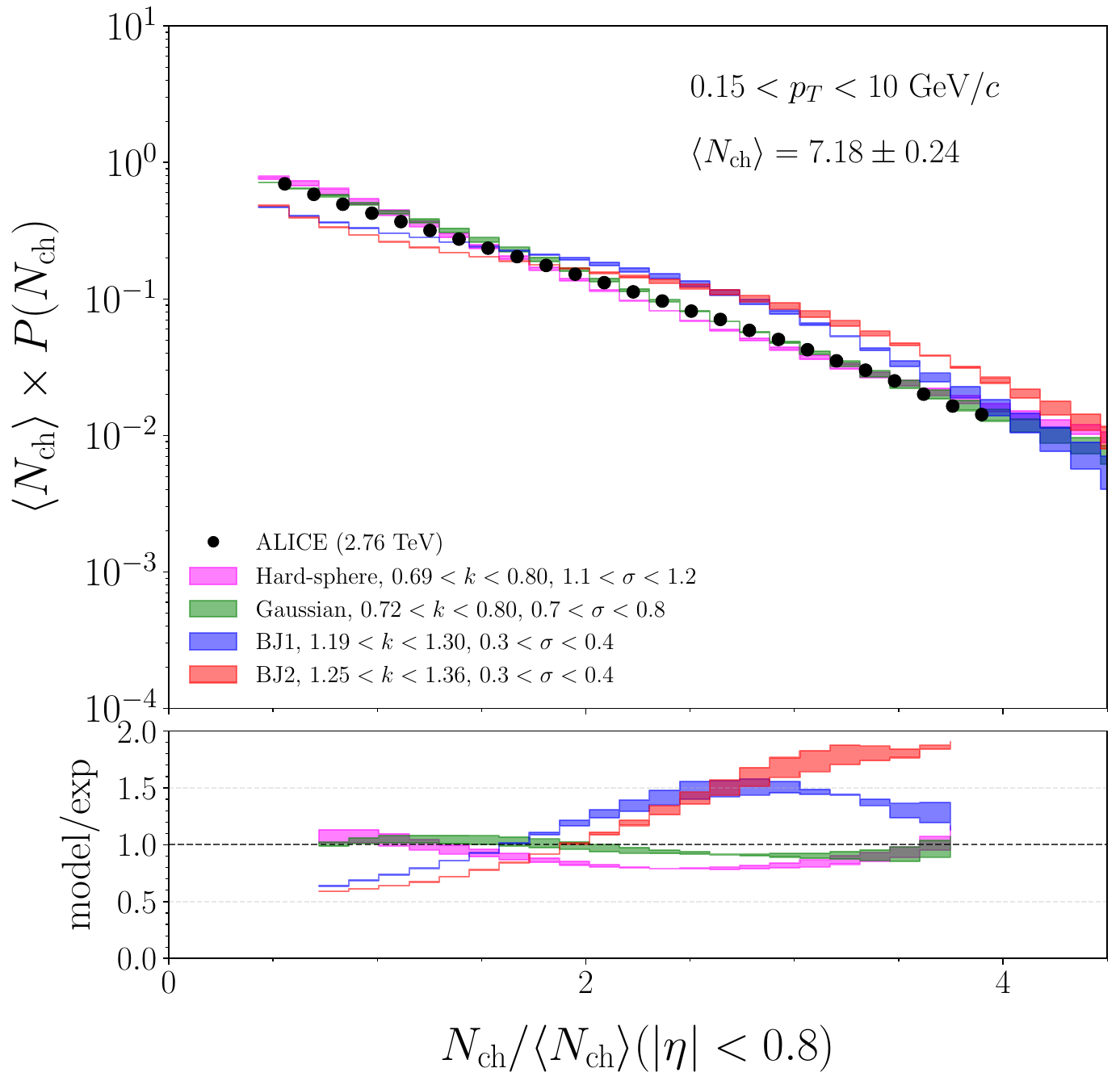}&
\includegraphics[scale=0.28]{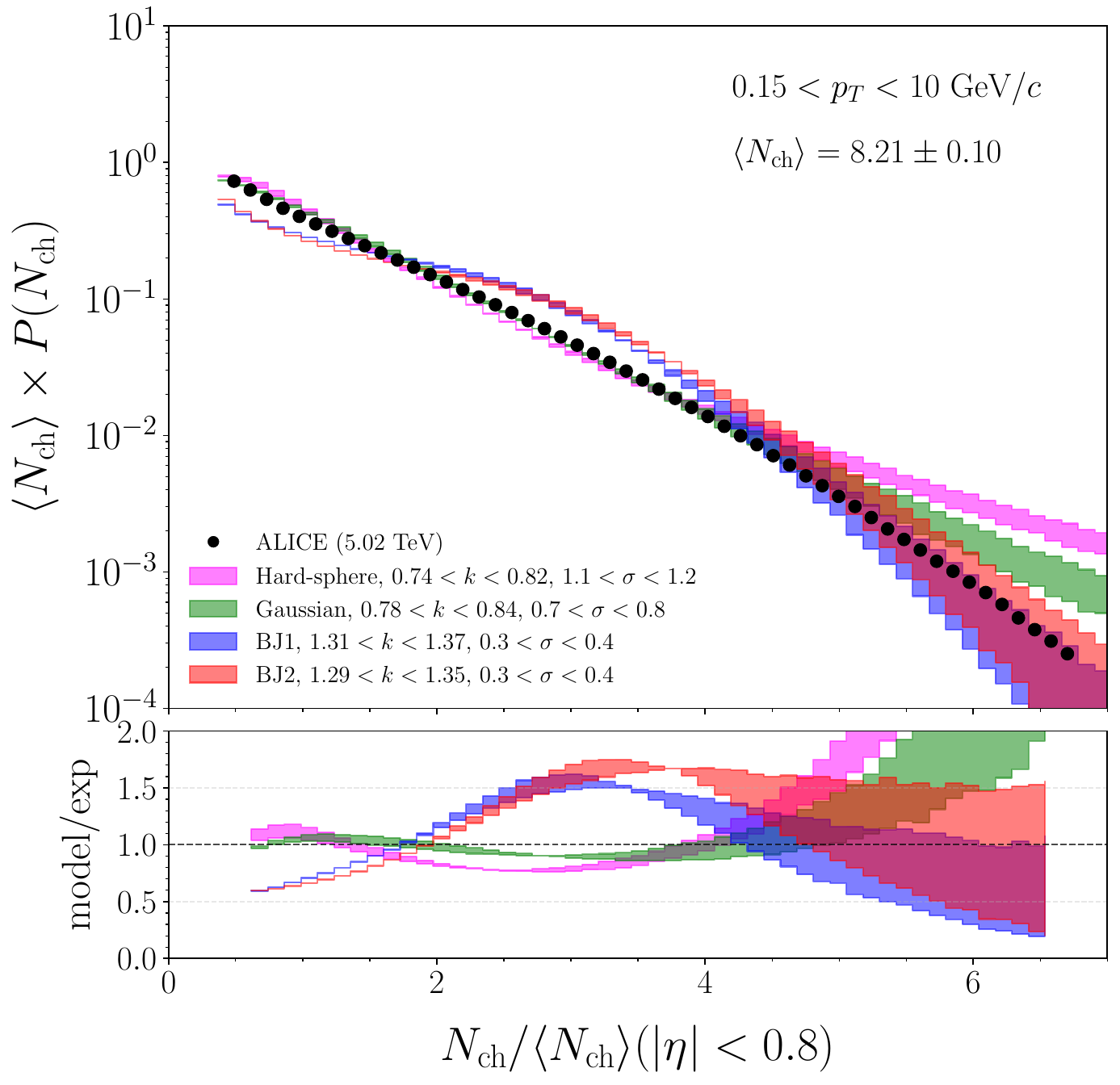} \\
(a) & (b)\\
\includegraphics[scale=0.28]{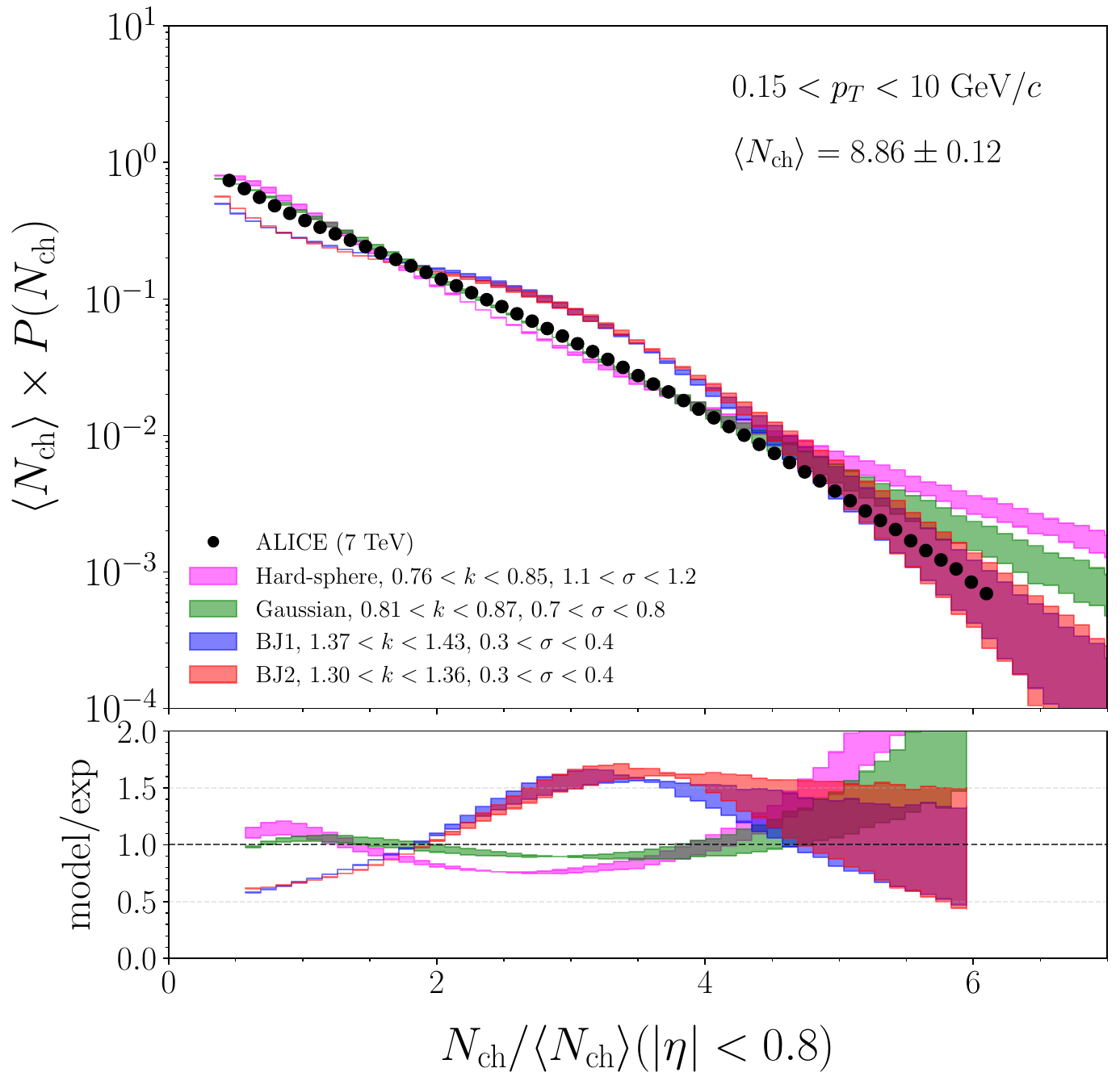}&
\includegraphics[scale=0.28]{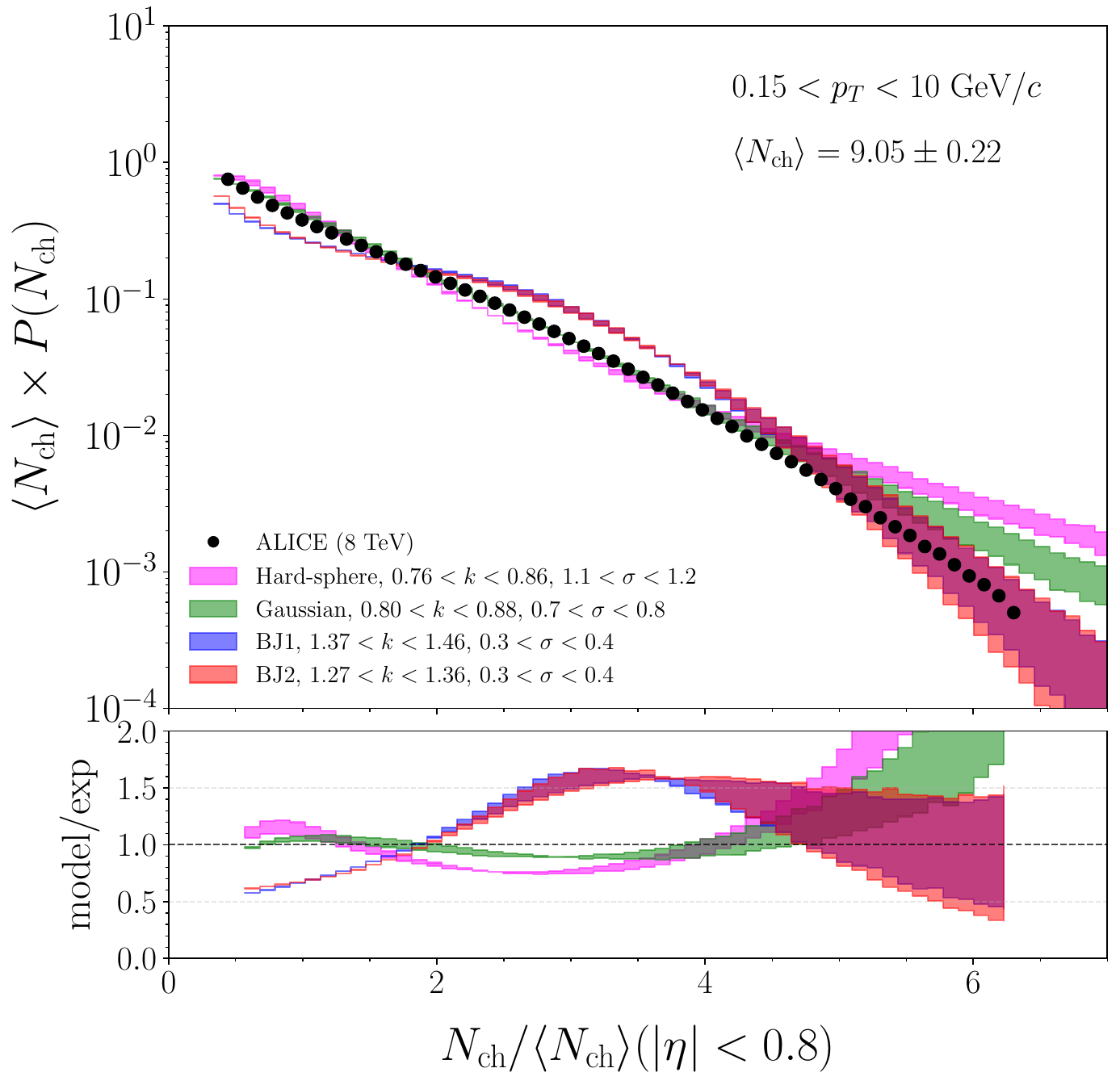} \\
(c) & (d) \\
\includegraphics[scale=0.28]{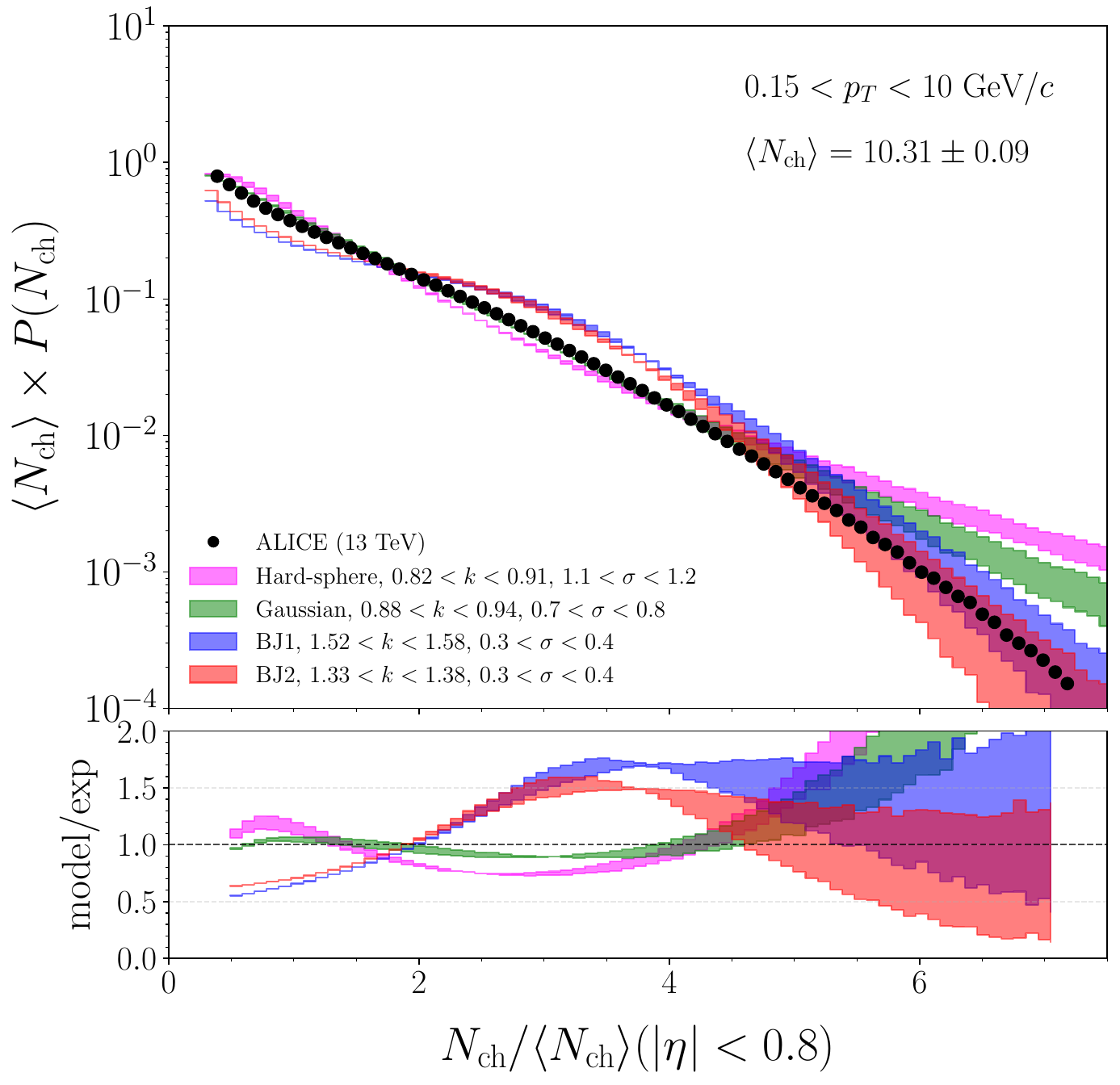}&
 \\
(e) & \\
\end{tabular}
\caption{KNO plot for pp multiplicity distribution                 
at a) $\sqrt{s}=2.76 \text{ TeV}$, b) $\sqrt{s}=5.02 \text{ TeV}$, 
c) $\sqrt{s}=7 \text{ TeV}$, d) $\sqrt{s}=8 \text{ TeV}$ and 
e) $\sqrt{s}=13 \text{ TeV}$. The four different lines correspond to the 
four different initial geometries. The experimental data 
(black points) are from \cite{alice24}.}
\label{fig3}
\end{figure}

\begin{figure}[h!]
\begin{tabular}{cc}
\includegraphics[scale=0.35]{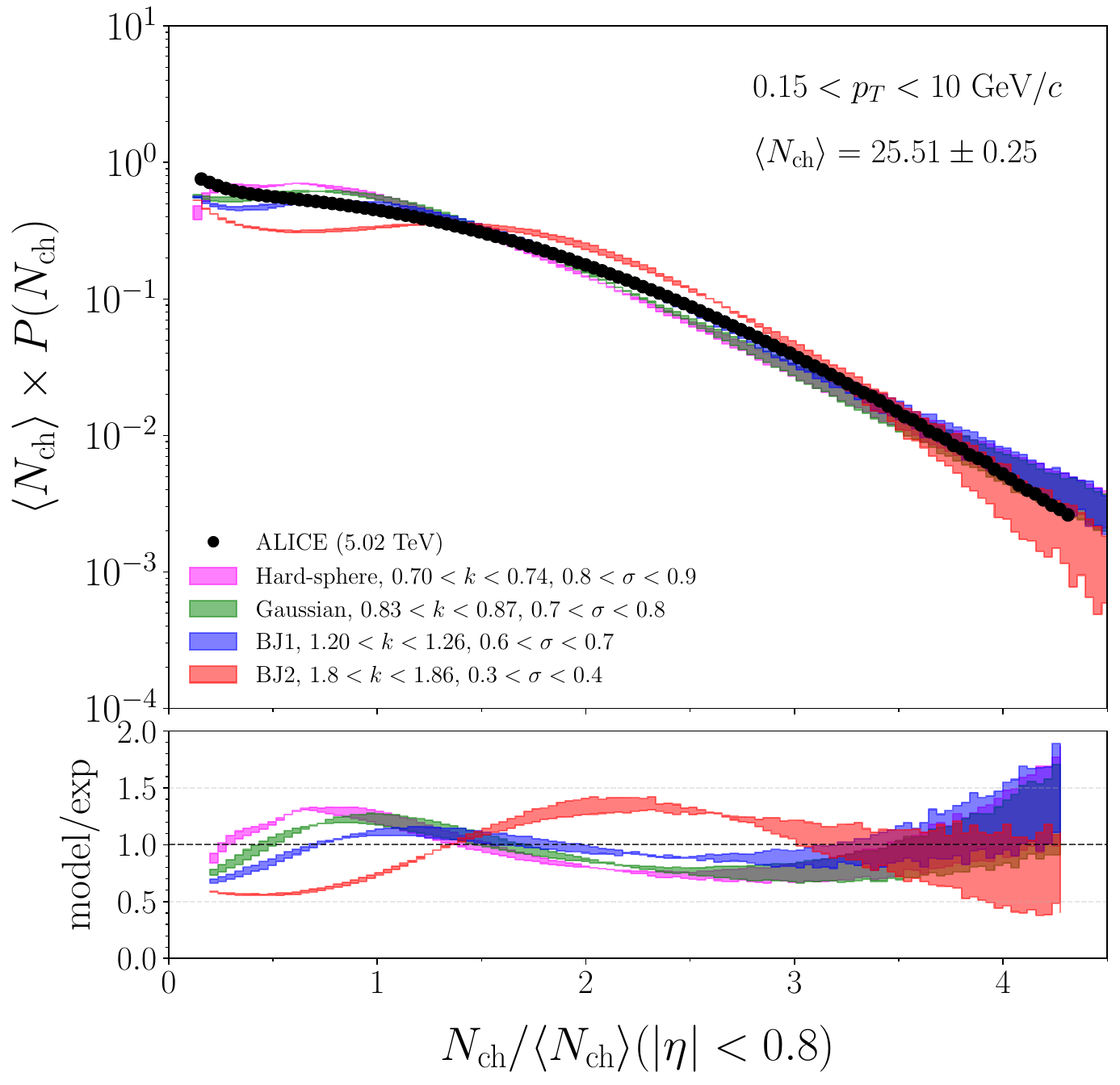}&
\includegraphics[scale=0.35]{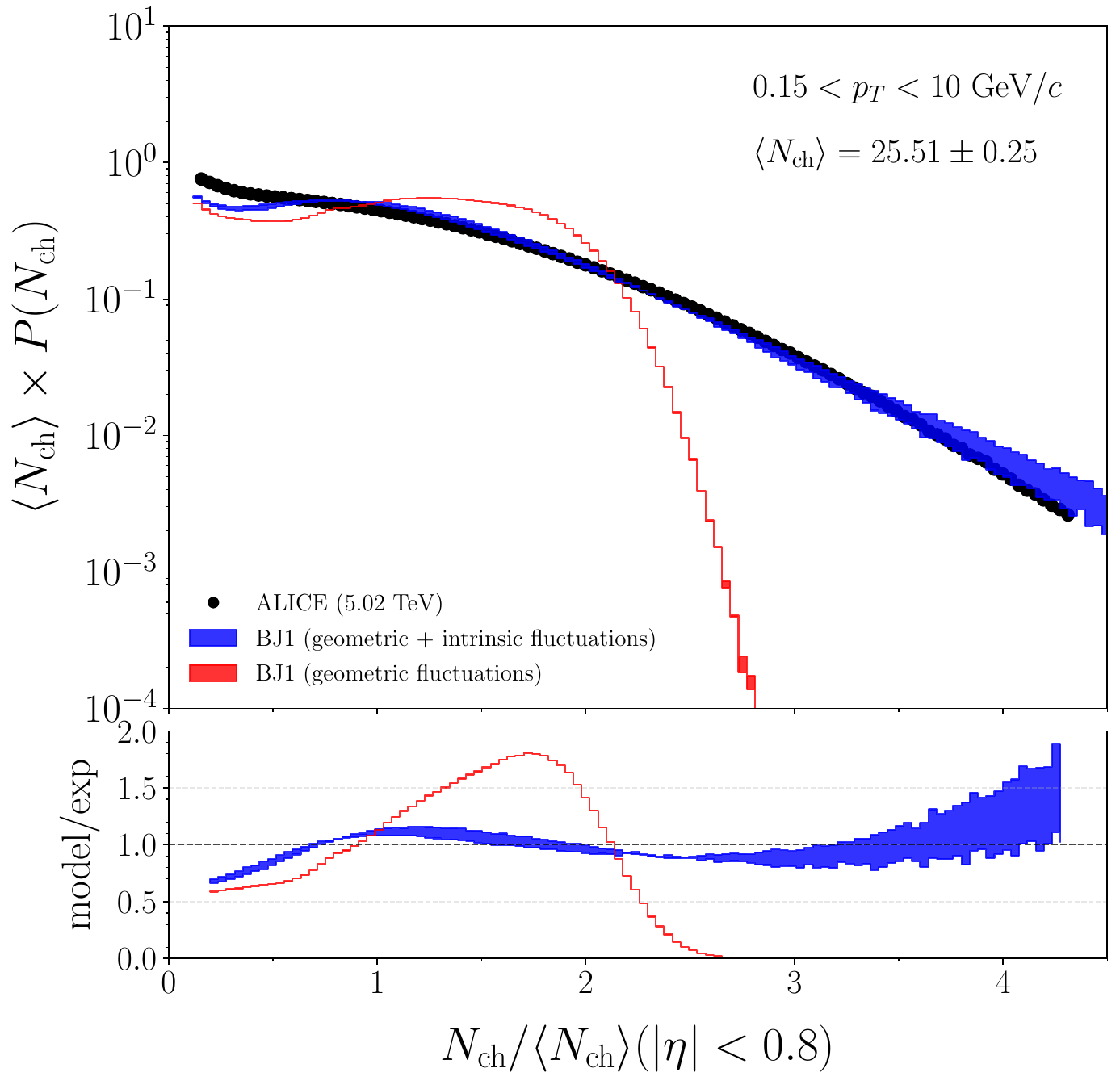} \\
(a) & (b)\\
\includegraphics[scale=0.35]{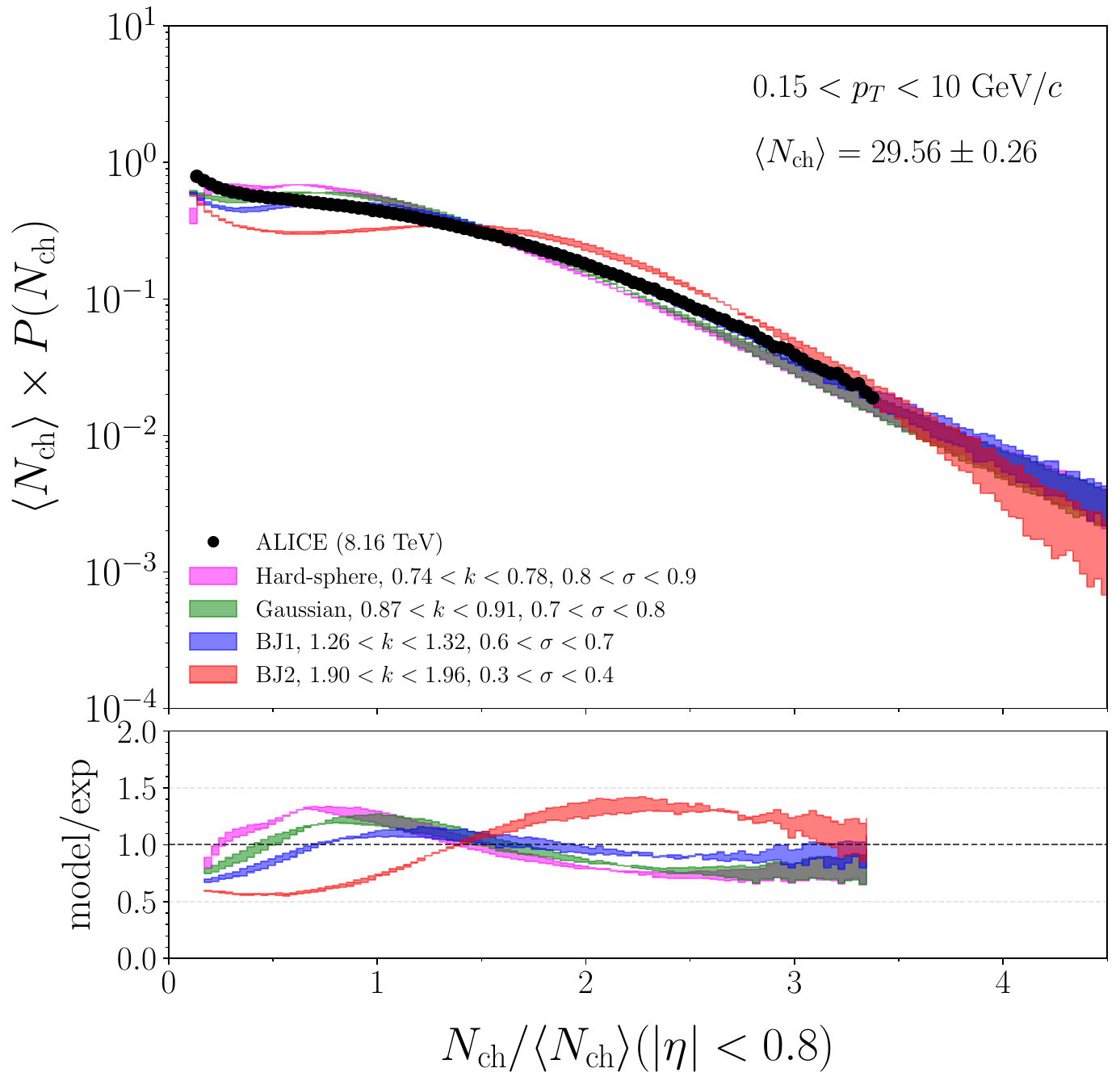}&
\includegraphics[scale=0.35]{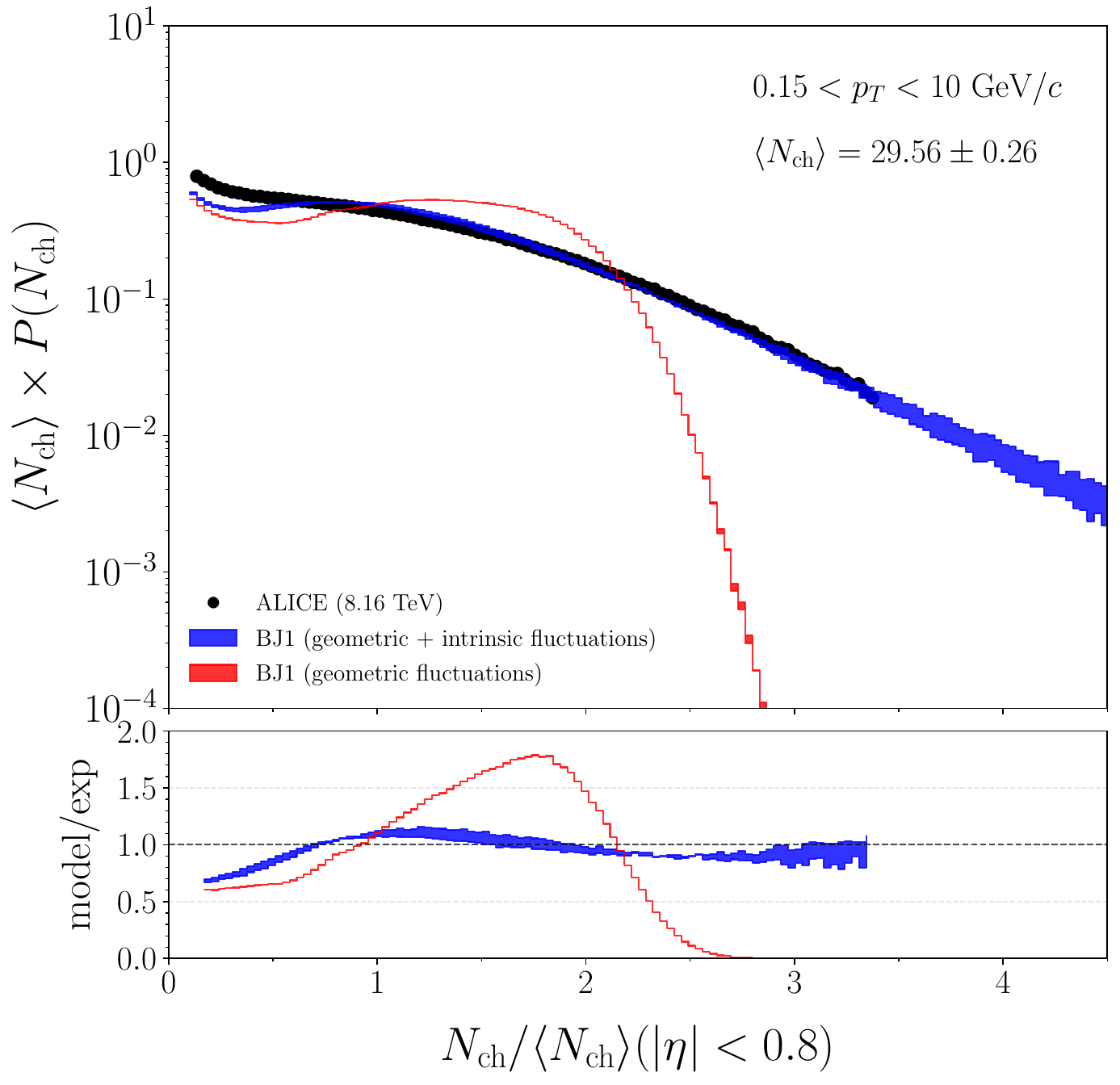} \\
(c) & (d)
\end{tabular}
\caption{KNO plot for pPb multiplicity distribution at             
a) $\sqrt{s}=5.02 \text{ TeV}$ and c) $\sqrt{s}=8.16 \text{ TeV}$. 
The four different lines correspond to the 
four different initial geometries. Comparison of BJ2 with and without 
intrinsic fluctuations at b) $\sqrt{s}=5.02 \text{ TeV}$ and 
d) $\sqrt{s}=8.16 \text{ TeV}$. The experimental data 
(black points) are from \cite{alice24}.}
\label{fig4}
\end{figure}

Integrating Eq.(\ref{ktfactorization}) in  $p_T$, in $y$ (or $\eta$) and in    
$r_{\perp}$ in an event-by-event basis, we construct the probability 
$P(N_{ch})$ of producing  $N_{ch}$ charged particles. The results for pp and 
pPb collisions are shown in Fig.\ref{fig3}  
and Fig.\ref{fig4}, respectively. In the figures $K$ is the multiplicative 
constant of  Eq.(\ref{ktfactorization}).  For fixed values of $K$ and $\sigma$ we
integrate the $k_T$ factorization formula and obtain values of $N_{ch}$. Accepting
all the values of $K$ and $\sigma$ which yield values of $N_{ch}$ which are
consistent with the experimental values (within the errors) and which are quoted in
the figures, leads to sets of values which define the areas  shown in the figures. 

The $k_T$ factorization formalism is appropriate to describe
inelastic non-diffractive events which yield  any number of produced particles
above a certain value, which we choose to be $N_{ch} = 3$. In terms of the KNO
variable, this means something like $N_{ch}/\langle N_{ch}\rangle \ge 0.5$.
Another limitation of our formalism is that  we did not include the contribution 
from the valence quarks.
They are supposed to be important only in the large rapidity regions, where
one can use the hybrid formalism \cite{duja}. At the  low rapidities considered here
their contribution should be less important but still needed for a better agreement 
with data.

\pagebreak

\pagebreak

Results for $P(N_{ch})$ are often presented in terms of the so called 
Koba-Nielsen-Olsen (KNO) scaling \cite{Koba:1972ng} variables. KNO scaling 
means that the multiplicity distribution,  $P(N_{ch})$, 
becomes energy independent at asymptotically high energies when we multiply it
by the mean value $\langle N_{ch}\rangle$  and plot it as a function of the 
scaling variable $z = N_{ch}/\langle N_{ch}\rangle$.  This feature has been experimentaly observed  at the LHC for narrow 
pseudo-rapidity ($|\eta|<0.5$) intervals \cite{CMS:2010qvf}; at the same 
time, the violation of KNO scaling has been reported for larger pseudo-rapidity 
intervals.

In order to reconcile the CGC formalism with KNO scaling, it has been shown 
that the theory must be nearly Gaussian (corresponding to the strong color field 
limit) and running coupling effects must be included as well \cite{Dumitru:2012tw}.
Such result was only verified for pp collisions up to $\sim 3.5$ times the 
average multiplicity. Small violations of the KNO scaling have been reported 
for pPb collisions in the region where $N_{ch} \lesssim 3$, where additional 
sources of fluctuations (geometric) are present \cite{Dumitru:2012yr}. 
Here, we extend the investigation to larger multiplicities. It is worth noting 
that KNO scaling might need to be modified for very large multiplicity 
events \cite{Moriggi:2025qfs}. In spite of all these efforts, a clear connection   
between KNO scaling and QCD is not yet established. For complemetary references on 
this subject we refer the reader to \cite{henrique,germano1,germano2}.

\begin{figure}[h!]
\begin{tabular}{cc}
\includegraphics[scale=0.35]{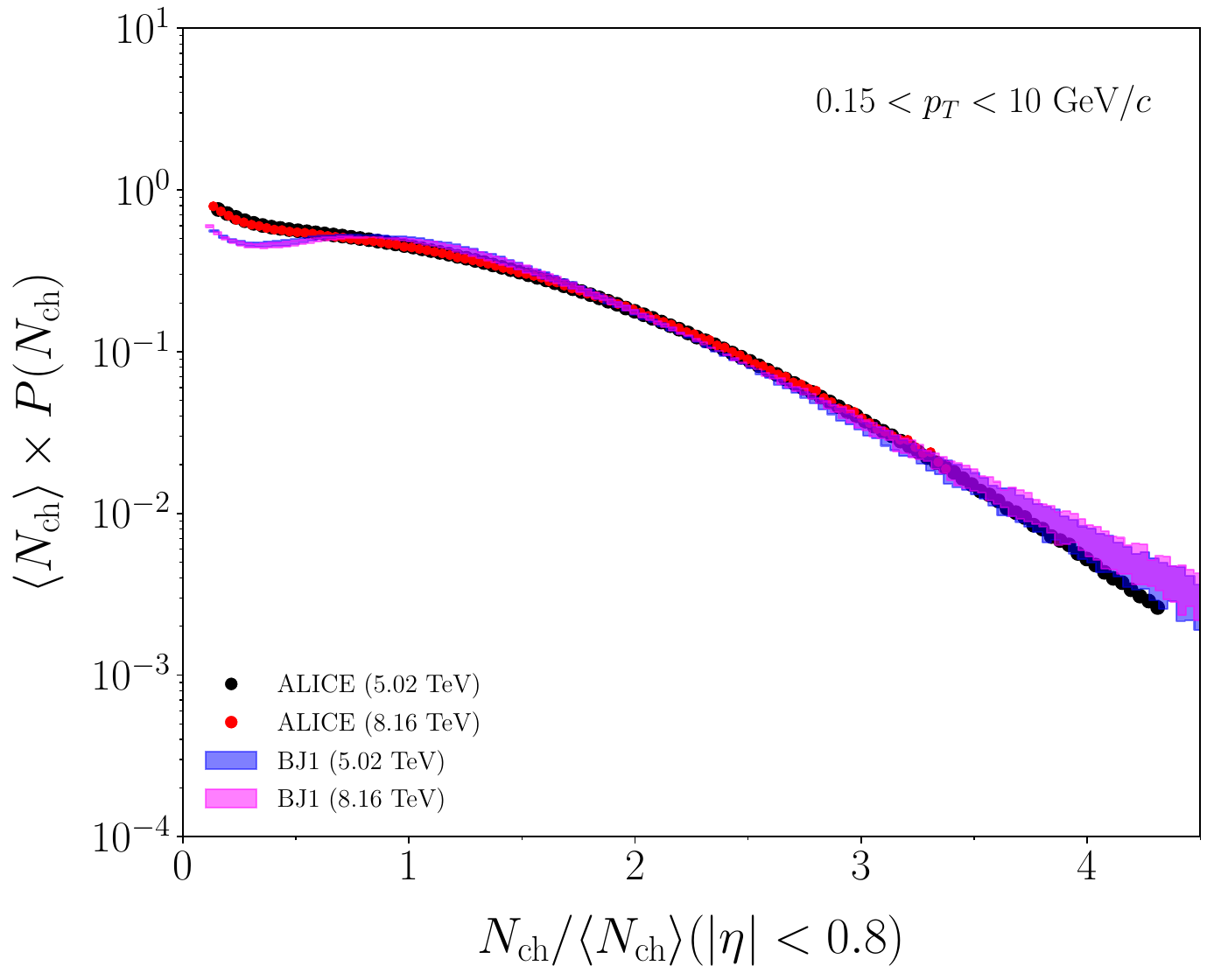} &                       
\includegraphics[scale=0.35]{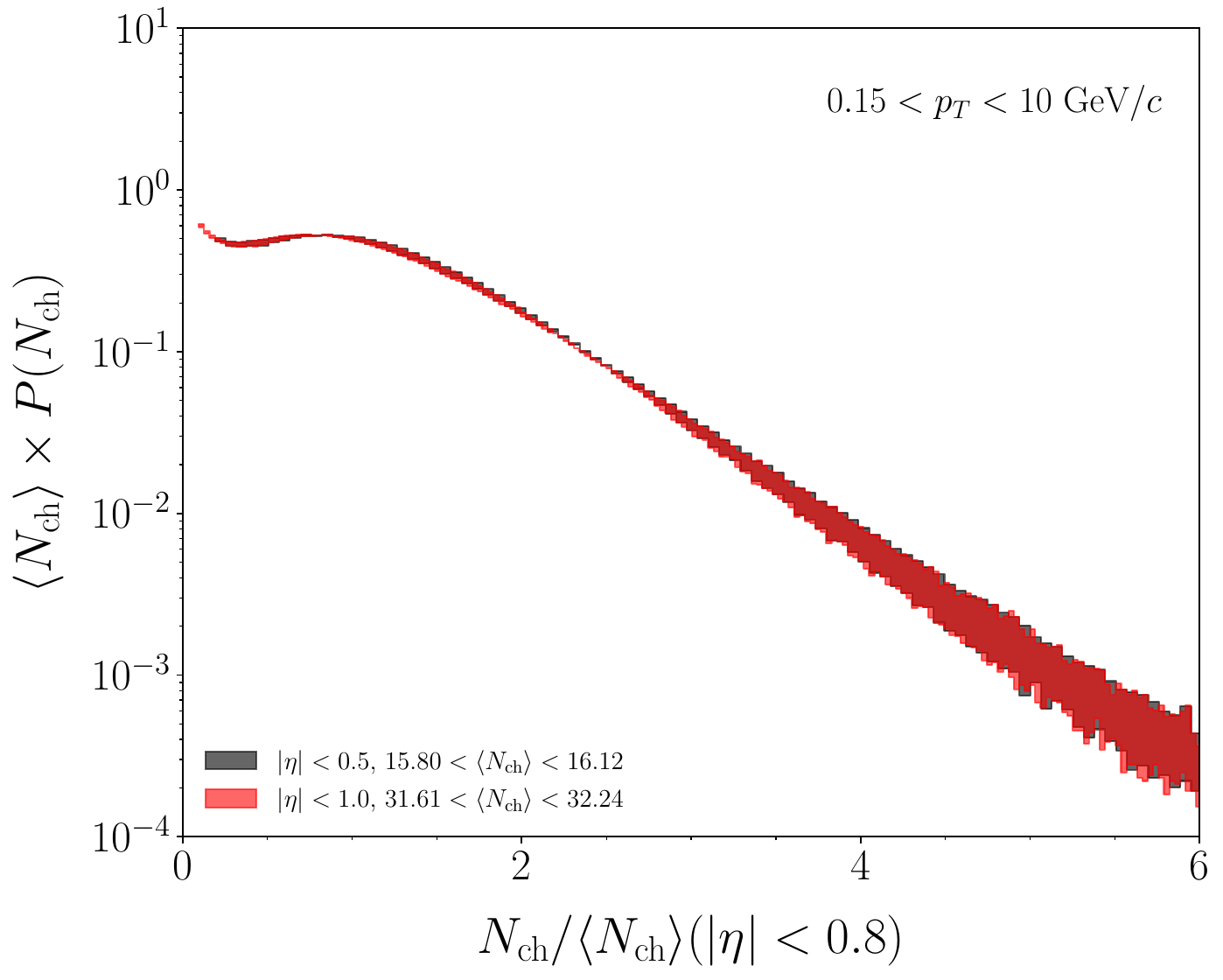}\\                                     
(a) & (b)\\ 
\end{tabular}
\caption{a) KNO plot for $\sqrt{s}=5.02$ and $8.16 \text{ TeV}$. 
The experimental data 
are from \cite{alice24}; b) Predictions for pPb multiplicity distributions  
at $\sqrt{s}=5.02 \text{ TeV}$ for two rapidity intervals $|\eta|<0.5$ and 
$|\eta|<1.0$ presented  in the KNO form.}
\label{fig5}
\end{figure}

The first measurements of  multiplicity distribution in pPb collisions 
\cite{alice24} were taken at two different energies. It is possible to make a
first check of the validity of the KNO, which, as can be seen in Fig.\ref{fig5} 
a, seems to be valid. In that figure we see that both experimental data and the 
MC-KLN model with BJ1 initial geometry exhibit KNO scaling.

Using the BJ1 initial condition we estimate the multiplicity distributions
for different pseudorapidity bins. In Fig.\ref{fig5} b we show our predictions
for  the pseudorapidity ranges $|\eta|<0.5$ and $|\eta|<1.0$. We also provide
the mean value estimate and observe that KNO scaling is still valid for these
pseudorapidity intervals.

Looking at Fig.{\ref{fig3}} we observe that data from  pp collisions  
are better described by Gaussian initial conditions, except for the region
$ z =  N_{ch}/  \langle N_{ch} \rangle \ge 4 $, where  BJ1 and BJ2 initial
conditions show a better agreement with the high multiplicity data. 
From Fig.\ref{fig4} we see that pPb data are  better described by BJ1. 
A remarkable feature is that the results from BJ1 and BJ2 are close to each other 
(within uncertainties) in pp collisions but are quite different in pPb collisions. 
In the low multiplicity region almost all curves slightly underpredict the data. 
In Figs.\ref{fig4}b and \ref{fig4}d we see the crucial role played by intrinsic
fluctuations at large values of $N_{ch}$.

To summarize, we would like to make the following remarks: first, from 
Figs.\ref{fig1} and ~\ref{fig2} we conclude that pPb collisions are better 
places to study the initial proton geometry than pp collisions; second, our model 
can be used to calculate the measured multiplicity distributions and the results 
are sensitive  to the initial state inputs. Further constraining the model 
parameters and reducing the theoretical error, the results would be different enough
to be discriminated by the data, which would ``prefer'' one initial condition. 
For the moment, in pPb and ultracentral pp collisions ($ z \ge 4 $) the data favor 
the baryon junction models. Moreover, in pPb collisions it is possible to 
discriminate between differents baryon junction models and BJ1 is clearly favored.
This is a consequence of the high sensitivity of pPb to the proton initial 
condition.

\section{Conclusion}

This work is part of a broader project of looking for experimental 
manifestations of the baryon junction configuration of the proton. Here we 
investigated proton-proton and  proton-nucleus collisions at the LHC. More     
specifically, we calculated the multiplicity distributions measured  in pp and 
pPb collisions. 
We used different initial state geometries as input in the MC-KLN Monte Carlo 
event generator which implements the $k_T$-factorization formalism of the CGC 
with KLN unintegrated gluon distributions. We found that in pp collisions the 
Gaussian spatial configuration is favored, except in the region $z \ge 4$, 
where BJ1 and BJ2 give a better description of the data. In pPb collisions the 
data are better explained by BJ1 initial conditions, provided that fluctuations 
in the saturation scale (also called ``intrinsic fluctuations'')  are included.

The existence of the baryon junction still needs confirmation. Further studies 
will be undertaken at the electron-ion collider.

\begin{acknowledgments}

We are grateful to M. Munhoz, H. Martins-Fontes, Y. Lima and M.V.T. Machado 
for useful discussions. This study was financed, in part, by the São Paulo 
Research Foundation       
(FAPESP), Brasil. Process Number 2024/06652-4. This work was financed by the 
Brazilian funding agencies CNPq, FAPESP, CAPES, FAPERJ and by the INCT-FNA. 
This work has been done as a part of the Project INCT-Física
Nuclear e Aplicações, Projeto No. 464898/2014-5.

The authors acknowledge the National Laboratory for Scientific Computing  
(LNCC/MCTI, Brazil), through the ambassador program (UFGD), subprojects 
FCNAE and SADFT for providing HPC resources of the SDumont supercomputer. 

\end{acknowledgments}

\end{document}